\documentclass{jfm}
\usepackage{graphicx}
\usepackage{epstopdf, epsfig}
\usepackage{amsmath}

\usepackage{color}
\usepackage[pdftex]{}

\shorttitle{Laminar-turbulent coexistence in annular Couette flow}
\shortauthor{K. Kunii, T. Ishida, Y. Duguet, and T. Tsukahara}

\title{Laminar-turbulent coexistence\\ in annular Couette flow}

\author{
Kohei Kunii\aff{1}, 
Takahiro Ishida\aff{1}\thanks{Present address: Japan Aerospace Exploration Agency, Tokyo 182-8522, Japan},
Yohann Duguet\aff{2},\\
\and 
Takahiro Tsukahara\aff{1}\corresp{\email{tsuka@rs.tus.ac.jp}}
}

\affiliation{
\aff{1}Department of Mechanical Engineering, Tokyo University of Science, Chiba 278-8510, Japan
\aff{2}LIMSI-CNRS, Universit\'e Paris-Sud, Universit\'e Paris-Saclay, F-91405 Orsay, France
}

\begin{document}

\maketitle

\begin{abstract}
Annular Couette flow is the flow between two coaxial cylinders driven by the axial translation of the inner cylinder. It is investigated using direct numerical simulation in long domains, with an emphasis on the laminar-turbulent coexistence regime found for marginally low values of the Reynolds number. Three distinct flow regimes are demonstrated as the radius ratio $\eta$ is decreased from 0.8 to 0.5 and finally to 0.1. The high-$\eta$ regime features helically-shaped turbulent patches coexisting with laminar flow, as in planar shear flows. The moderate-$\eta$ regime does not feature any marked laminar-turbulent coexistence. In an effort to discard confinement effects, proper patterning is however recovered by artificially extending the azimuthal span beyond 2$\pi$. Eventually, the low-$\eta$ regime features localised turbulent structures different from the puffs commonly encountered in transitional pipe flow. In this new coexistence regime, turbulent fluctuations are surprisingly short-ranged. {Implications are discussed in terms of phase transition and critical scaling}. 
\end{abstract}



\section{Introduction}


Most wall-bounded shear flows possess a linearly stable base flow for parameters where turbulence can also be observed. This is particularly true
near the onset of turbulence, i.e. near the smallest Reynolds number $Re_g$, `g' for global, where turbulence is sustained. There is no contradiction between linear stability of the laminar base flow
and the existence of a turbulent regime, since the latter can be reached from well-chosen finite-amplitude perturbations to the base flow \citep{Orszag71, Schmid01}.
Such turbulent flows are {called} subcritical, and there is no general method to identify the value of $Re_g$. A large number of studies have demonstrated that  turbulent regimes near $Re_g$
are intermittent in space and time. At every instant in time they consist of trains of localised coherent structures {such as} turbulent puffs in pipe flow \citep{Wygnanski73, Wygnanski75, Bandyopadhyay86, Barkley15} and turbulent stripes in planar shear flows such as plane Couette or plane Poiseuille flow \citep{Prigent02,Barkley05,Tsukahara05,Tsukahara10,Fukudome12,Seki12}.

A turbulent puff is a turbulent region localised in the axial direction of the pipe. A mechanism for the self-sustenance of this coherent structure has been identified by \cite{Shimizu09}. {However puffs do not persist indefinitely}. As demonstrated by \cite{Hof06}, puffs {have finite lifetimes} as long as they keep their localisation. The onset Reynolds number $Re_g$ is determined for the case of pipe flow as the value of $Re$ at which their decay rate and proliferation rate exactly balance each other \citep{Avila11}, i.e. $2040 \pm 10$ when $Re$ is based on the flow rate and the pipe diameter.

Turbulent stripes are large-scale coherent structures {emerging spontaneously} near $Re_g$ in a variety of planar wall-bounded flows such as {plane Couette flow (pCf)} with or without rotation \citep{Barkley07,Tsukahara10,Duguet10,Tsukahara2010RPCF,Brethouwer12}. They display obliqueness with respect to the mean flow direction, explained by the robust presence of oblique large-scale flows  near the laminar-turbulent interfaces \citep{Duguet13}. The onset Reynolds number of pCf is estimated as $Re_g \approx 324 \pm 1$ using the classical definition of $Re$, from either experimental lifetime measurements \citep{Bottin98} or numerical decay experiments \citep{Duguet10}. {Such structures were initially reported in counter-rotating regimes of the Taylor-Couette flow (TCf) {\citep{Coles65,Prigent02}} and are not specific to perfectly planar geometries.} 

Whereas the coherent structures and the associated mean flows near $Re_g$ for pipe flow can be described as  one-dimensional, patterning in pCf is more genuinely two-dimensional. Establishing {universal links} between all the different sustained coherent structures (puffs, stripes) in such flows is an ambitious task. By noting that these coherent structures are (turbulent) steady states, {one can borrow from bifurcation theories the concept of \emph{continuation}, i.e. a continuous deformation from one flow case to another one, hoping that there is indeed at least one path in parameter space which continuously {links} one type of solution (here stripes) to the other one (puffs)}. If such a path exists then puffs and stripes can be considered members of a unique family of subcritical flows. {Continuation methods have shown to be useful for the tracking of nonlinear steady states {from one model problem to another}. The first such instance was considered by \cite{Nagata90} who deformed tertiary solutions of TCf into non-trivial steady states of pCf.} In the same spirit, \cite{faisst2000transition} suggested a criterion to assess {whether and when} the curvature of TCf can be considered geometrically negligible.

\begin{figure}
\begin{center}
\includegraphics[width=0.7\linewidth]{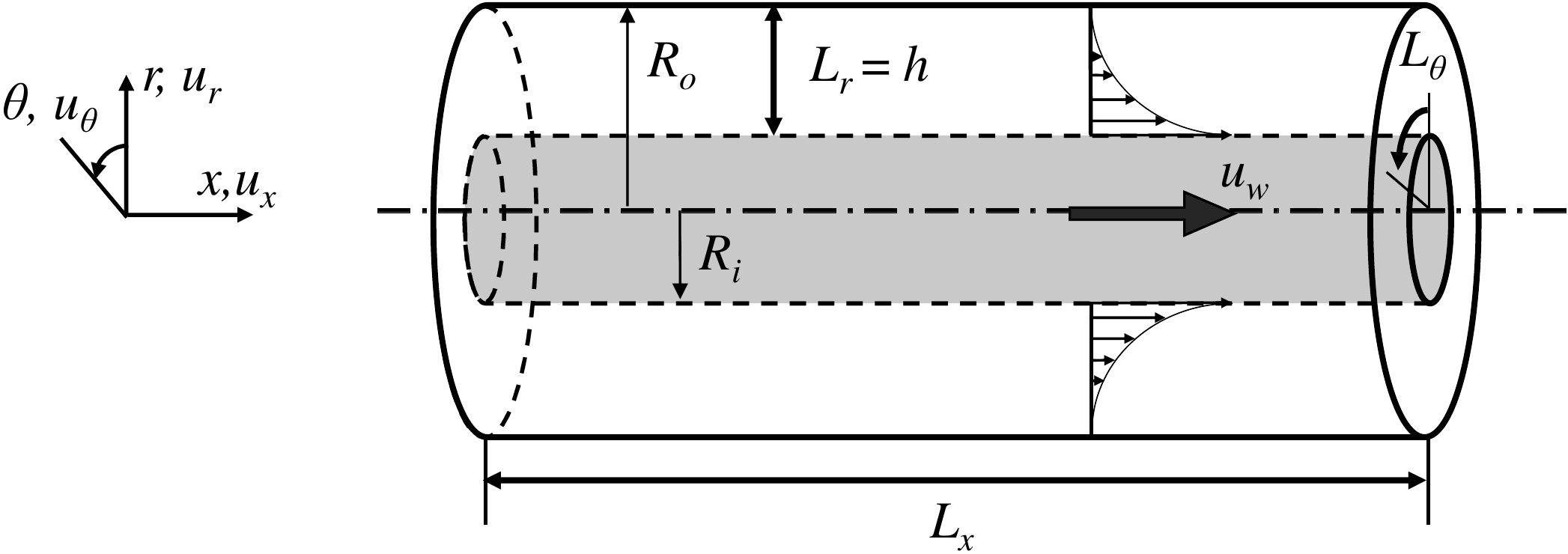}
\end{center}
\caption{Configuration of annular Couette flow, with notations.}
\label{fig:scf_model}
\end{figure}

\begin{figure}
\begin{center}
\includegraphics[height=5cm]{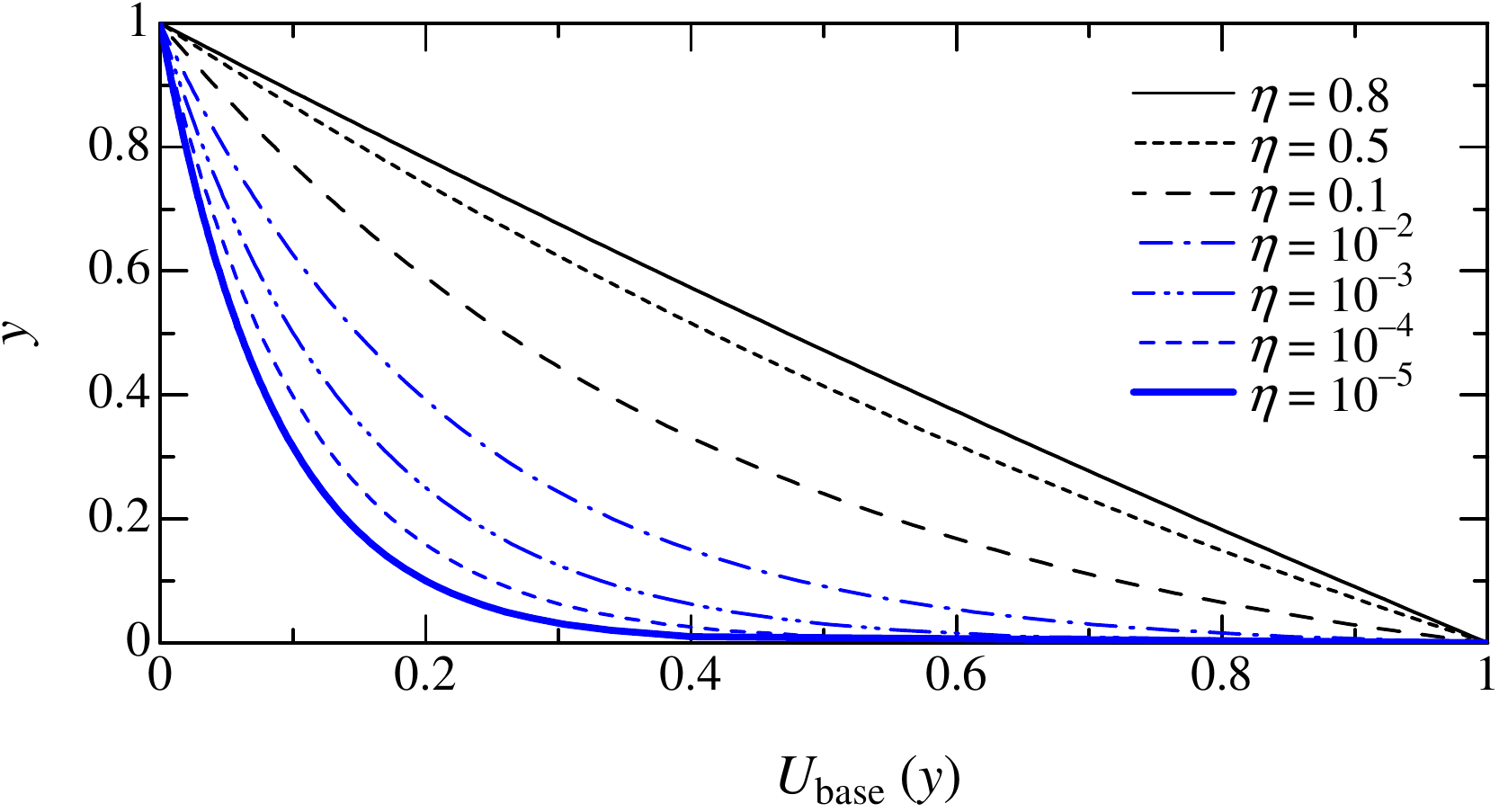}
\end{center}
\caption{Laminar velocity profiles of annular Couette flow at several radius ratios. The values of $\eta$ investigated in this article ($\eta=0.8$, 0.5, and 0.1) are the three top-most curves (black online), while smaller values of $\eta$ down to $10^{-5}$ are displayed using a clearer font (blue online). {See \S \ref{sec:equation} for the definition of $y$.}}
\label{fig:ubase}
\end{figure}

In this article, we focus on a new candidate flow for continuation between pipe flow puffs and plane Couette stripes : the annular Couette flow (aCf). It refers to the flow between two infinitely long coaxial cylinders, driven by the axial translation of the inner rod with constant velocity $u_w>0$. A sketch is displayed in figure~\ref{fig:scf_model}. 
{This flow is an idealised academic configuration inspired by thread-annular flows \citep{Frei00,Walton03,Walton04,Walton05,Webber08}, for which large-scale coherent structures can limit the degree of homogeneity of the processed material}. The main non-dimensional geometric parameter in aCf is the radius ratio $\eta = R_i/R_o$ (where $R_i$ and $R_o$ denote respectively the inner radius and outer radius). The radius difference $h=R_o-R_i$, represents the gap between the two cylinders. As $\eta \rightarrow 1$, the wall curvature vanishes and the flow is asymptotically equivalent to pCf. {At the other end, for $\eta \rightarrow 0$, the geometry does not become a cylindrical pipe because of the presence of the inner rod (associated with no slip). There is hence no possible continuation from aCf to pipe flow. Besides the difference in topology, the velocity distributions of pipe flow and aCf in the vanishing $\eta$ limit strongly differ. This is readily seen by analysing {the analytical laminar solutions} shown in figure \ref{fig:ubase}. For $\eta \rightarrow 1$ the profile clearly tends uniformly towards the centrosymmetric linear profile of pCf, whereas for 
$\eta \rightarrow 0$ the wall shear is located at the inner rod only, 
unlike pipe flow for which the shear is located at the outer {boundary}. A similar conclusion is expected for turbulent regimes but deserves a full quantitative study.} The Reynolds number is defined as $Re=u_wh/4\nu$, where $\nu$ is the kinematic viscosity of the fluid. That definition is consistent with the pCf limit as $\eta$ approaches 1. 

A number of stability investigations have focused on the linear stability of the base flow as a function of $\eta$ and $Re$. According to \cite{Gittler93}, aCf is linearly stable against axisymmetric perturbations for $\eta > 0.1415$,
while it becomes unstable at {finite but very large} Reynolds numbers ($\Rey > 10^6$) for $\eta < 0.1415$. \cite{Liu12} demonstrated the possibility for large non-modal growth of perturbations in the linearly stable regimes for most values of $\eta$ and for values of $Re$ much below the linear stability threshold. By considering non-axisymmetric perturbations and tracking numerically exact nonlinear travelling waves, \cite{Deguchi11} determined {the} values of $Re_g$ of 255.4, 256.6, and 288.6 for $\eta$ = 1$^{-}$, 0.5, and 0.1, respectively. To our knowledge, no study has tracked turbulent coherent structures for fixed $\eta$ all the way towards their extinction at $Re_g$.

\begin{figure}
\begin{center}
(a)\includegraphics[width=0.7\linewidth]{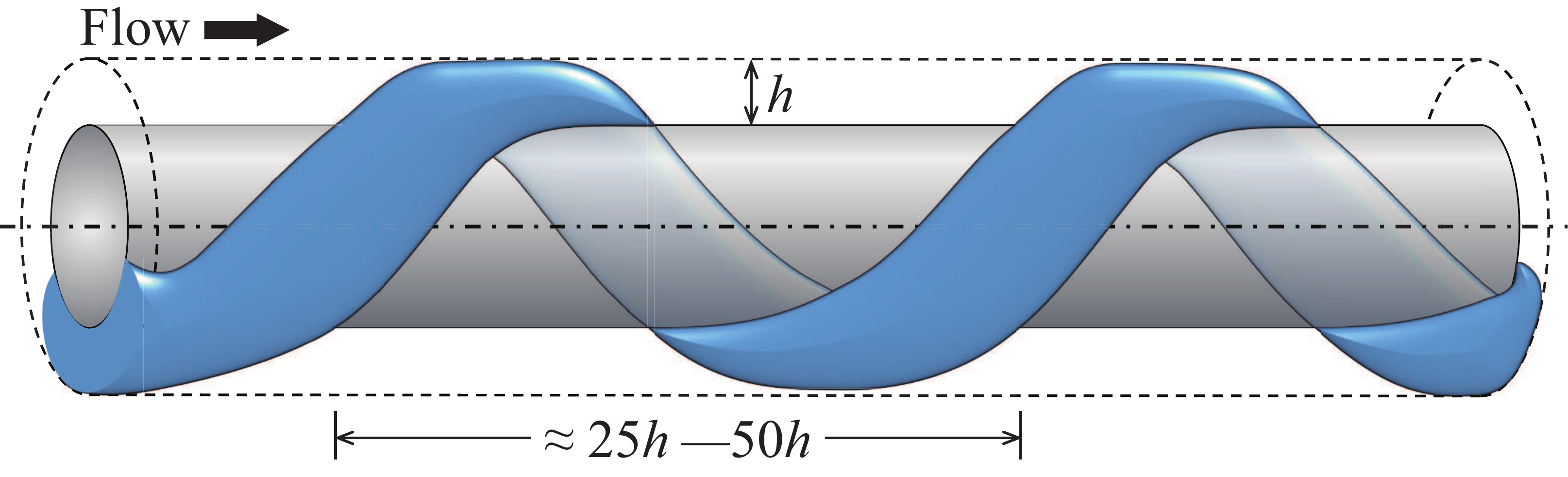}
\\ \vspace{0.5em}
(b)\includegraphics[width=0.7\linewidth]{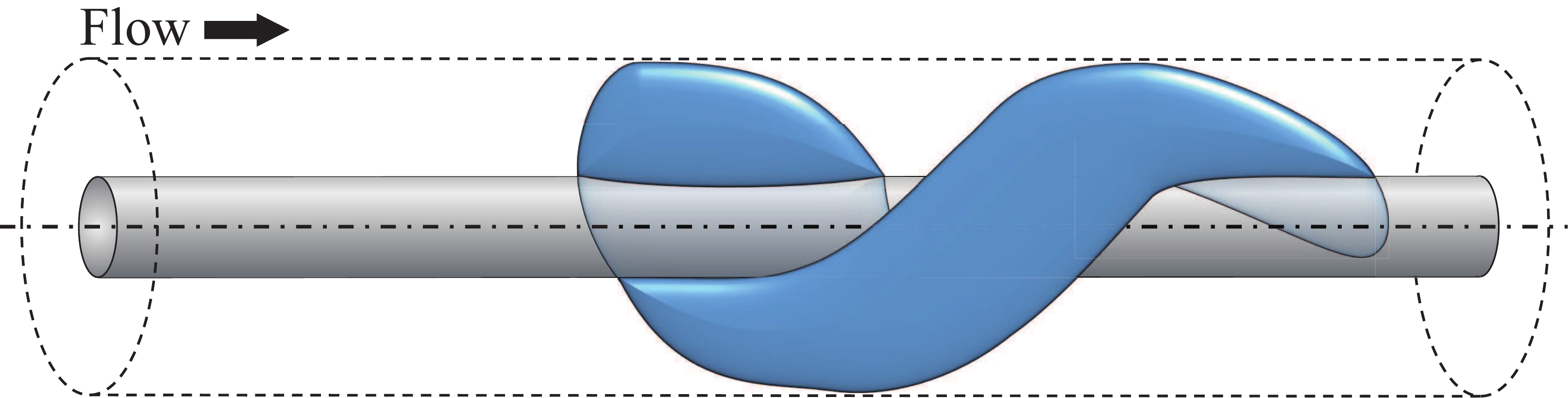}
\\ \vspace{0.5em}
(c)\includegraphics[width=0.7\linewidth]{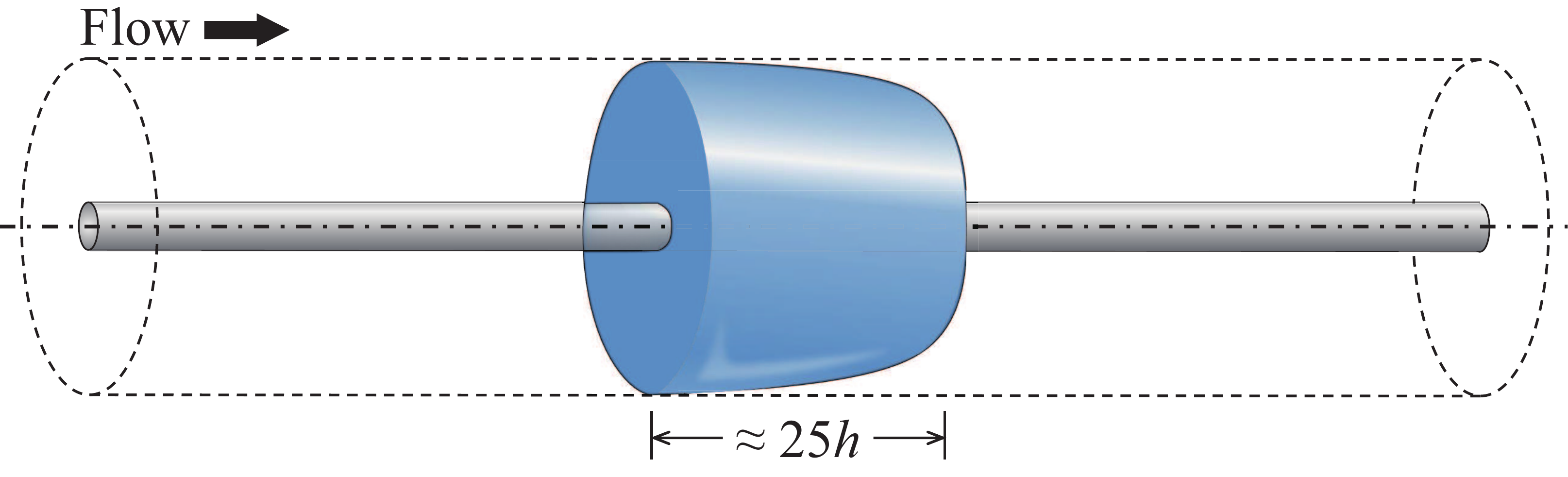}
\end{center}
\caption{Typical transitional structures in annular flow geometries depending on $\eta$, according to \cite{Ishida16}.
Blue regions represent localised turbulence. The aspect ratio (gap \emph{vs} streamwise length) is changed for each figure.
(a) helical turbulence for high $\eta$, (b) helical puff for intermediate $\eta$, and (c) straight puff for low $\eta$.}
\label{fig:tran_st}
\end{figure}

 It is the goal of the present paper to describe the distinct representative regimes of localised turbulence in aCf near $Re_g$. The mechanisms that can take aCf away from pCf for decreasing $\eta$ are twofold : either azimuthal confinement prevents large-scale coherent structures from forming, or the wall curvature is such that the coherent structures themselves change form. These two mechanisms, though both linked to finite $\eta$, are different. We take inspiration from a cousin of aCf, namely aPf (annular Poiseuille flow), the flow occurring in the same geometry, yet with a non-moving inner rod and a pressure gradient applied against the {axial} direction. The different transitional regimes of aPf were described by \cite{Ishida16}. Three types of coherent structures were identified depending on the value of $\eta$, all sketched in figure~\ref{fig:tran_st} : helical stripes for $\eta$ close to but below unity, straight puffs for low enough $\eta$, and an intermediate state occurring near $\eta \approx 0.3$ in the form of {a} helical puff, i.e. a helical stripe of finite streamwise extent.  \cite{Ishida17} performed a statistical analysis of the transition from puffs to stripes as $\eta$ is varied, by investigating statistically the large-scale flows present at the laminar-turbulent interfaces. They revealed a statistical cross-over from straight puffs (figure~\ref{fig:tran_st}c) to oblique laminar-turbulent interfaces {(figure~\ref{fig:tran_st}b)}, for radius ratios $\eta \approx 0.3$--0.4, {where the flow consists} of a spatially disordered mixture of helical and straight puffs. The main mechanism breaking the helical stripes of aPf is essentially azimuthal confinement : by noting that the lengthscale of interest is the annular gap width $h$, and {by} expressing all lengths in units of {$h$}, ``large-scale" flows can only be accommodated {inside} the gap for $\eta$ close enough to unity. The common geometry to aCf and aPf leads {to the expectation that} transitional structures occurring in aCf are similar to those found in aPf. To this end we have chosen to focus on the transitional regimes occurring for the {same} values of $\eta=$ 0.8, 0.5 and 0.1. For each case, we report whether {or not} turbulent stripes form {as} in pCf. In the absence of stripes, the mechanical reason for it, either {azimuthal} confinement or wall curvature, will be analysed. As will be described throughout the paper, the planar picture is valid for large $\eta \approx 1$, where the flow {stays} similar to pCf, but it breaks down for {lower} $\eta \le 0.5$.
 
The present paper is structured as follows. Section~\ref{sec:method} contains the governing equations and the description of the numerical tools. In \S~3, the intermittent regimes are described for three different values of $\eta=0.8$, 0.5, and 0.1 in decreasing order, with a focus on whether the flow patterns are influenced by {the} azimuthal confinement or not. Section~\ref{sec:statistics} contains temporal statistics of the flow field as well as measures of intermittency. Finally the universal features of transitional aCf are discussed in the concluding \S~5.

\section{Governing equations and numerical methods}
\label{sec:method}

\subsection{Governing equations}
\label{sec:equation}

We consider an incompressible Newtonian fluid flow between two coaxial cylinders. The outer cylinder of radius $R_o$ is fixed whereas the inner cylinder of radius $R_i$  moves in the axial direction $x$ with velocity $u_w>0$.
The gap has size $h = R_o - R_i$. The quantities $h$ and $u_w$ are used to non-dimensionalise the governing equations. The flow configuration is {sketched} in figure~\ref{fig:scf_model}.
The non-dimensional coordinates $r$ and $\theta$ are, where necessary, converted respectively into $y=r-r_i$ and $z=r\theta$ where $r_i \le r \le r_o$, with $r_i=R_i/h$ and $r_o=R_o/h$. The non-dimensional velocity field ${\bf u} \equiv (u_x, u_r, u_\theta)$ in cylindrical coordinates and the pressure field are governed by the incompressible Navier--Stokes equations :
  \begin{equation}
   \nabla \cdot {\bf u} = 0,
   \label{eq:continuity}
  \end{equation}
  \begin{equation}
    \frac{\partial {\bf u}}{\partial t} 
    + ( {\bf u} \cdot \nabla ) {\bf u}
    = - \nabla p
    + \frac{1}{\Rey} \nabla^2 {\bf u},
   \label{eq:ns_x}
  \end{equation}
where $p$ stands for the reduced pressure field. {The fluid density is unity}. \\

The Reynolds number is defined as $\Rey \equiv u_w h /4\nu$, with $\nu$ the kinematic viscosity of the fluid. 
The total wall shear stress is linked to the friction Reynolds number $\Rey_\tau$, defined as 
  \begin{equation}
   \Rey_\tau = \frac{u_{\tau}h}{2\nu},
   \label{eq:retau}
  \end{equation}
  It is based on the friction velocity $u_{\tau}$ evaluated
at each wall using the formula
 \begin{equation}
   u_{\tau} = \frac{\eta u_{\tau,\mathrm{inner}}+u_{\tau,\mathrm{outer}}}{\eta +1},
   \label{eq:utauavg}
  \end{equation}
where each $u_{\tau}$ is defined by the corresponding wall shear stress $\tau_w$ and by the relation $\tau_w=\rho u_{\tau}^2$, from which inner units can be defined (see table \ref{tab:condition}).\\

The boundary conditions are periodicity in the streamwise $(x)$ and the azimuthal coordinates ($\theta$), and no-slip at the walls : 
  \begin{eqnarray}
    {\bf u}=(1,0,0) & {\rm at} & r=r_i~({\rm or}~y=0), \label{eq:bc1}\\
    {\bf u}=(0,0,0) & {\rm at} & r=r_o~({\rm or}~y=1), \\
    {\bf u}(0,r,\theta) & = & {\bf u}(L_x,r,\theta), \\
    {\bf u}(x,r,0) & = & {\bf u}(x,r,2n\pi).  \label{eq:bc}
  \end{eqnarray}
{In the last boundary condition (\ref{eq:bc}) expressing the azimuthal periodicity, the value of $n$ is usually $n=1$. Choosing $n$ as {a fraction} of the form $1/m$, with $m$ a positive integer, would correspond to {imposing} an additional discrete azimuthal symmetry. In \S~3, we will take the original choice of considering $n$ as a positive integer such as $n=1,2,3,...$. Such a boundary condition leads to an azimuthal coordinate ranging from 0 to $2\pi n$, which for $n>1$ lacks direct physical interpretation. This is {intended} here as a numerical trick in order to question the effect of confinement by the boundary conditions.}\\

{The governing equations (\ref{eq:continuity})--(\ref{eq:ns_x}) together with the} boundary conditions (\ref{eq:bc1})--(\ref{eq:bc}) admit a steady axial solution of the form
  \begin{equation}
  U_{\rm base}(y)= \frac{\ln{\left[y(1-\eta)+\eta \right]}}{\ln{\eta}},
  \label{eq:base}
  \end{equation}
interpreted as the laminar base flow. For $\eta \rightarrow 1$, $U_{\rm base}$ converges in a non-singular way towards the linear profile $U_{\rm base}(y)|_{\eta=1}\sim (1-y)$ which is consistent with the pCf limit.
{Figure~\ref{fig:ubase} {displays} profiles of $U_{\rm base}(y)$ parameterised by $\eta$ ranging down to $10^{-5}$, for which the profile unambiguously differs from that in cylindrical pipe flow.}
 
\subsection{Numerical tools}

Equation~(\ref{eq:ns_x}) is discretised in space using finite differences. The time discretisation is carried out using a second-order Crank--Nicolson scheme, and an Adams--Bashforth scheme for the wall-normal viscous term and the other terms{, respectively}. Further details about the numerical methods used here can be found in \cite{Abe01}.

\begin{table}
\begin{center}
\begin{tabular}{cccc}
$\eta$                           & 0.1  & 0.5  & 0.8 \\\hline
$\Rey$                    & 750 $\rightarrow$ 390 & 750 $\rightarrow$ 377.5 & 750 $\rightarrow$ 325 \\
$\Rey_\tau$               & 34.5 $\rightarrow$ 14.7 & 49.0 $\rightarrow$ 18.8 & 51.3 $\rightarrow$ 18.1 \\
$R_i$, $R_o$                     & 0.111$h$, 1.11$h$ & $h$, 2$h$  & 4$h$, 5$h$ \\
$L_x \times L_r \times L_\theta$ & $204.8h \times h \times 2\pi$ & $102.4h \times h \times 2\pi$ & $102.4h \times h \times 2\pi$ \\
$L_{zi}$, $L_{zo}$               & 0.698$h$, 6.98$h$  & 6.28$h$, 12.6$h$  & 25.1$h$, 31.4$h$ \\
$N_x \times N_r \times N_\theta$ & $1024 \times 64 \times 256$ & $1024 \times 64 \times 256$ & $1024 \times 64 \times 512$ \\
{$\Delta x$} & {$0.2h$} & {$0.1h$} & {$0.1h$} \\
{$\Delta r$}  & {$0.00132h$--$0.0316h$} & {$0.00132h$--$0.0316h$} & {$0.00132h$--$0.0316h$} \\
{$\Delta z_i$, $\Delta z_o$}  & {$0.00273h$, $0.0273h$} & {$0.0245h$, $0.0491h$} & {$0.0491h$, $0.0614h$}  \\
$\Delta x^+$                     & 2.94--6.90 & 1.88--4.92 & 1.81--5.13 \\
$\Delta r^+$                     & 0.0387--1.09 & 0.0495--1.56 & 0.0477--1.62 \\
$\Delta z^+$                     & 0.0400--0.940 & 0.461--2.42 & 0.888--3.15 \\
\end{tabular}
\end{center}
\caption{Computational conditions for aCf with nominal azimuthal extent $L_{\theta}=2\pi$ : radius ratio $\eta$, Reynolds number $\Rey$, friction Reynolds number $\Rey_\tau$, inner radius $R_i$, outer radius $R_o$, computational domain dimensions {$L_x \times L_r \times L_\theta$}, inner perimeter $L_{zi}$, outer perimeter $L_{zo}$, number of grid points $N_x \times N_r \times N_\theta$, and range of values for the increments $\Delta j$ ($j = x, r, \theta$ (or $z$)). The superscript ($^+$) denotes the normalisation by wall units based on $u_\tau$ and $\nu$.}
\label{tab:condition}
\end{table}

The tested radius ratios are 0.8, 0.5, and 0.1, and the other numerical parameters are shown in table \ref{tab:condition}.
The number of grid points in the azimuthal direction is 256 for $\eta = 0.1$ and 0.5 with the non-uniform radial mesh, but we used 512 grid points for $\eta = 0.8$ because of its wider domain relative to the gap $h$.
The strategy to isolate the transitional regimes of aCf is similar to that used in aPf. {First,} using an arbitrary finite-amplitude 
initial condition, a turbulent flow is computed at large enough $\Rey$ for which no laminar-turbulent coexistence is expected.
In a second phase, $\Rey$ is decreased in small steps until the flow can be considered statistically steady based on energy time series. Visual inspection of the flow fields at mid-gap is used to decide whether or not the flow displays laminar-turbulent coexistence.

\section{Morphology of coherent structures}

\subsection{High $\eta=0.8$}

\begin{figure}
\begin{center}
(a)\includegraphics[width=0.95\linewidth]{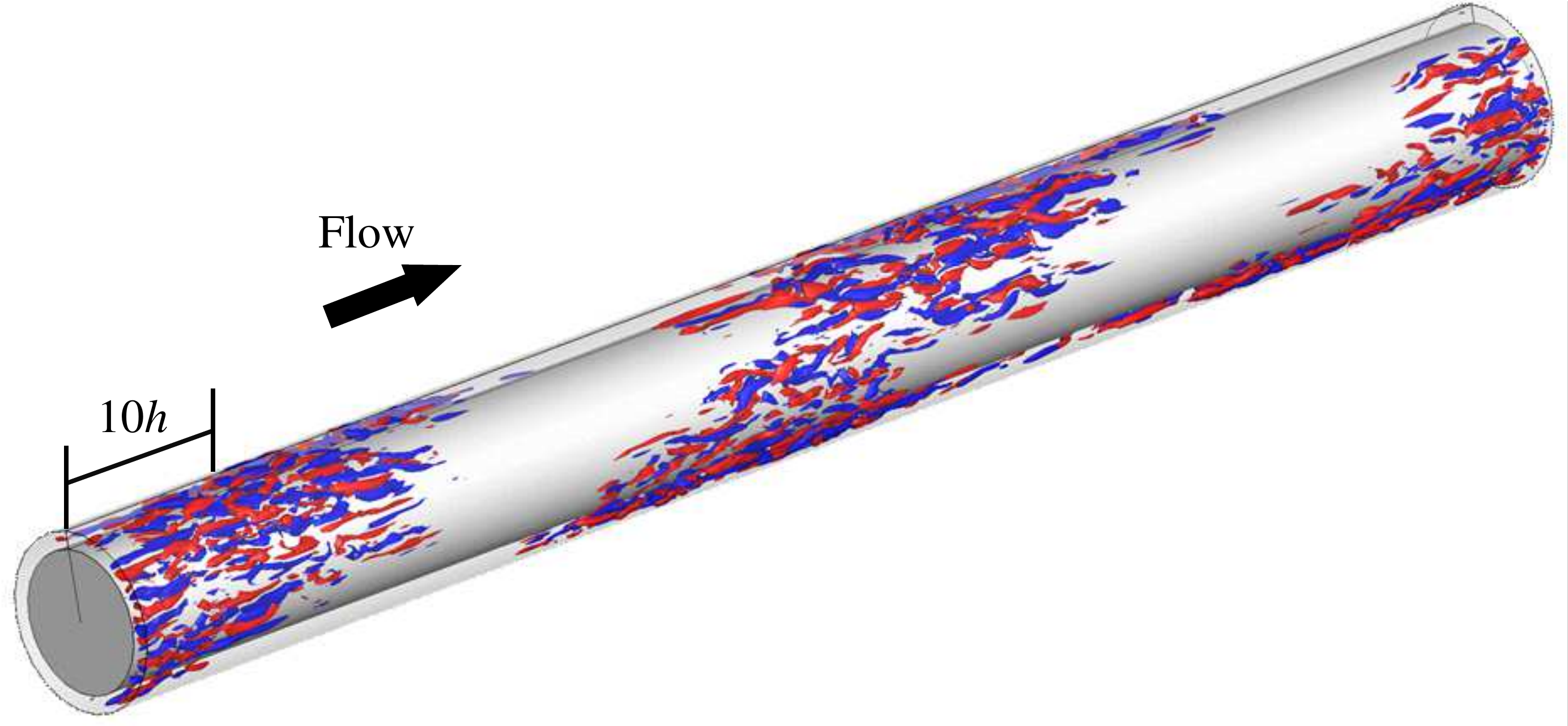} \\ \vspace{0.5em}
(b)\includegraphics[width=0.95\linewidth]{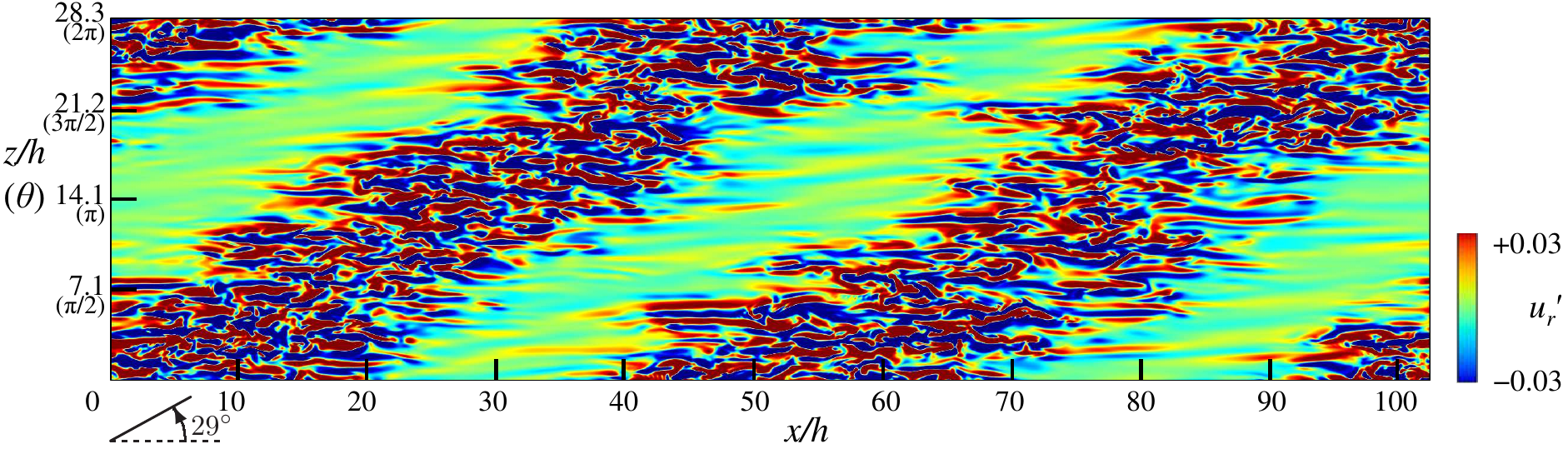} \\ \vspace{0.5em}
\end{center}
\caption{Instantaneous flow field of aCf for $\eta = 0.8$ and $\Rey = 350$, accompanied by helical turbulence. (a) Three-dimensional visualisation by iso-surfaces of wall-normal velocity fluctuations ($u_r^\prime =-0.03$, blue; $u_r^\prime =0.03$, red); (b) two-dimensional contour of $u_r^\prime$ in the $x$-$z$ (or $x$-$\theta$) plane at mid-gap. The pattern propagates steadily towards $x>0$ (to the right in figure).}
\label{fig:tran08}
\end{figure}

We begin the exploration of the intermittent regimes by considering a relatively high value of $\eta=0.8$, for which the curvature is weak and aCf is expected to behave like plane Couette flow (pCf).
Figure~\ref{fig:tran08} shows the wall-normal velocity of a typical flow field for $\eta = 0.8$ and $Re=350$.
The unfolded plane of two-dimensional contours {in figure~\ref{fig:tran08}b} lies at mid-gap ($y=0.5$) {where the corresponding spanwise width} is $28.3h$. Measurements of the friction Reynolds number yield $\Rey_\tau = 25.2$.
Expressed in wall units, the domain size visualised in figure~\ref{fig:tran08}b is equivalent to $L_x^+ \times L_z^+ = 5160 \times 1430$.
As expected, the large-scale spatial patterning of the flow field is evident whereas at smaller scales {streamwise streaks dominate}. The turbulent region wraps around the inner cylinder in the form of a helix, whose pitch angle is approximately $29^{\circ}$ with respect to the $x$ axis. The laminar streamwise intervals have lengths of about $50h$, consistently with those in pCf \citep{Prigent02, Duguet10, Brethouwer12}. Such helical turbulence also occurs in aPf \citep{Ishida16} with a similar value of the pitch angle.
By reducing $\Rey$, the helical turbulence sustains until $Re = 337.5$ ($Re_\tau = 24.0$) and eventually decays at 325. This scenario and the estimated value for $Re_g$ are fully consistent with those for pCf, despite the curvature of the walls.

\subsection{Moderate $\eta=0.5$}

\subsubsection{Nominal parameters}

Let us now focus on the moderate $\eta$ regime by analysing the case $\eta=0.5$. Note that in the context of aPf, the transitional regime for $\eta=0.5$ differs little from the case $\eta=0.8$. The analysis in \cite{Ishida17} shows that for $\eta > \eta_c \approx 0.3$, the perimeter is sufficiently large (in units of $h$) for large-scale flows to form, and as a consequence localised turbulence can only take the shape of helical puffs as in figure~\ref{fig:tran_st}b or helical stripes for even higher $\eta$. At these marginal Reynolds numbers, the smallest vortical structures occupy the whole gap and $h$ is hence the convenient reference lengthscale. {The fact that $h$ defines the order of magnitude for the size of the smallest lengthscales, has been suggested as the condition defining marginality \citep{Alfredsson2000}.}

\begin{figure}
\begin{center}
(a)\includegraphics[width=0.95\linewidth]{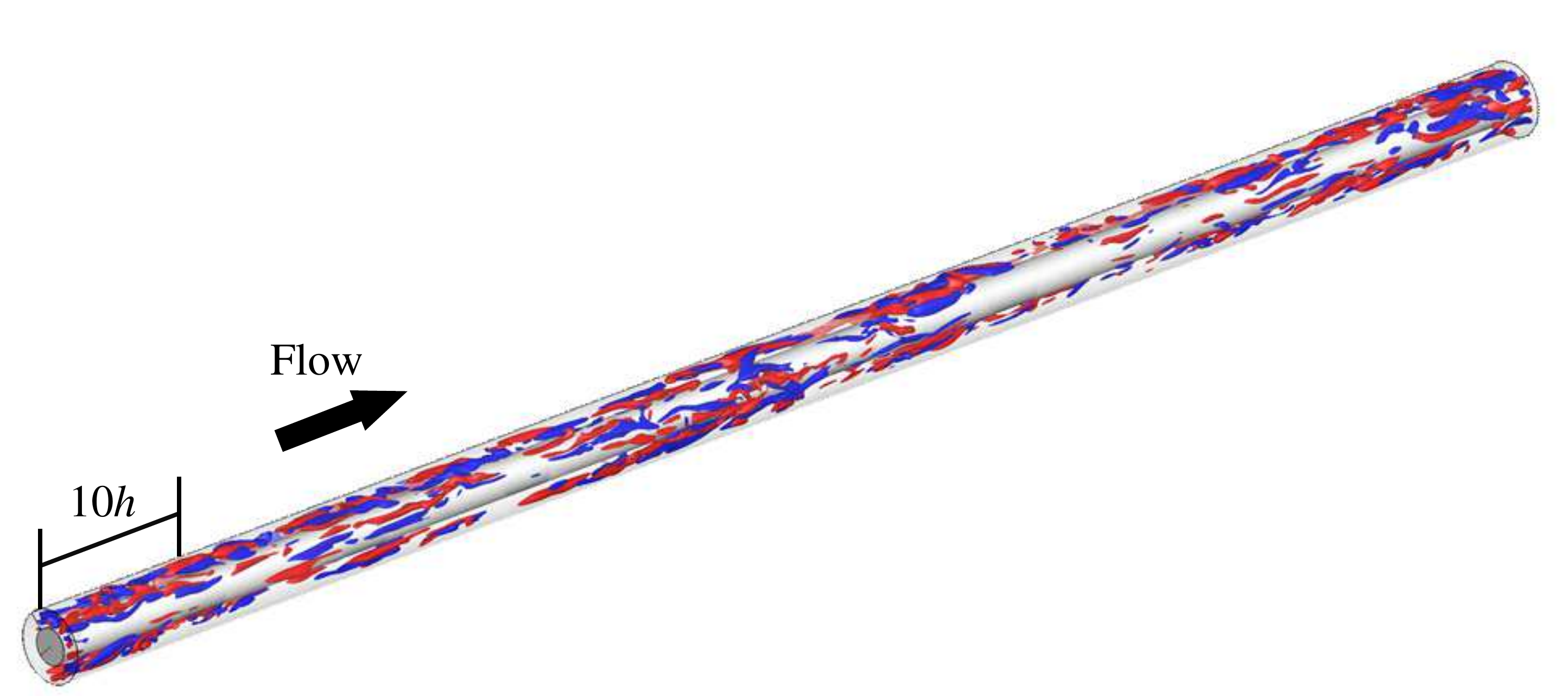} \\ \vspace{0.5em}
(b)\includegraphics[width=0.95\linewidth]{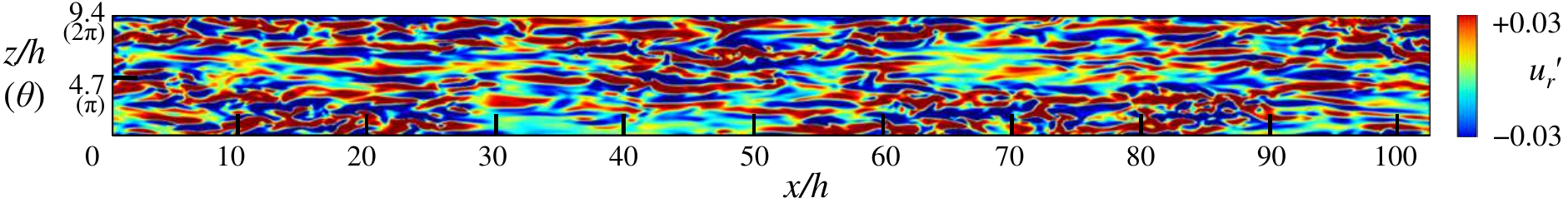} \\ \vspace{0.5em}
\end{center}
\caption{Same as figure~\ref{fig:tran08} for $\eta = 0.5$ and $\Rey = 387.5$, featuring laminar patches.}
\label{fig:tran05}
\end{figure}

Whereas helical stripes prevail for $\eta=0.8$, they are absent from the simulations at $\eta=0.5$. A typical flow regime obtained for $\Rey = 387.5$ is displayed in figure~\ref{fig:tran05}. In this case, the azimuthal length is only $9.4h$ at mid-gap, which is shorter than the (intrinsic) laminar interval of turbulent bands.
As a result, despite a furtive occurrence, proper helical turbulent bands cannot sustain as organised structures. Some disorganised laminar patches can be observed in practice near $\Rey = 380$ ($\Rey_\tau = 27.2$).
Decreasing $\Rey$ further, the turbulent regions become intermittent in the streamwise direction but no band-like or puff-like structure forms.
The flow eventually relaminarises at $\Rey = 377.5$, a value unambiguously higher than the threshold value for $\eta = 0.8$.

\subsubsection{Azimuthally extended system}
\label{sec:8pi}

As mentioned in the introduction, two intermingled factors can influence the formation of large-scale structures in comparison with
the planar case : the azimuthal extent and the wall curvature. We use a numerical trick to test whether the obtained flows
persist in the absence of azimuthal confinement \emph{without} modifying the wall curvature, namely by simulating the flow in a domain 
where the azimuthal variable describes the range $(0, 2n\pi)$ with $n>1$. The hypothesis $L_{\theta}>2\pi$ should be seen as a deliberate validation technique rather than as the introduction of an esoteric parameter. 

{The strategy to seek transitional regimes featuring laminar subdomains is similar to that for $L_{\theta}=2\pi$.} For the case $\eta=0.5$, we have tried several integer values of $n$ until visual inspection of the flow fields reveals some clear change{s with respect to $n=1$}. The case $n=4$ corresponds to a periphery of $L_{\theta}=8\pi$, i.e. 4 nominal peripheries, and the perimeter values at the inner wall, at mid-gap and at the outer wall are $(L_{zi}, L_{zc}, L_{zo}) = (25.1h, 37.6h, 50.2h)$, respectively.
The number of {azimuthal grid points} is increased to $N_\theta = 512${, hence doubled in resolution with respect to $n=1$.} In other words there was a trade-off in the numerical resolution, resulting probably in a quantitative shift of the Reynolds numbers.
The extended azimuthal grid spacings on the outer and inner walls are, respectively, $\Delta z = 0.098h$ and $0.049h$, which remain adequate resolutions when expressed in inner units since they verify  $\Delta z^+ \le 5$.

\begin{figure}
\begin{center}
(a)\includegraphics[width=0.95\linewidth]{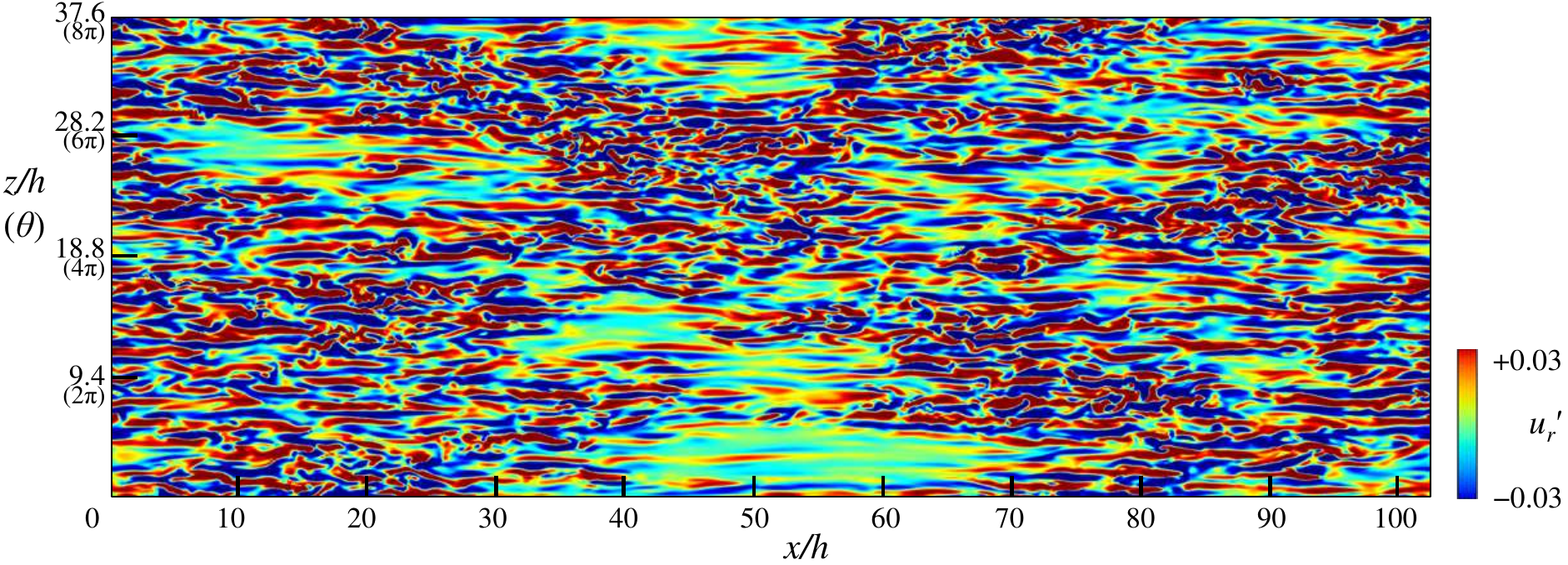} \\ \vspace{0.5em}
(b)\includegraphics[width=0.95\linewidth]{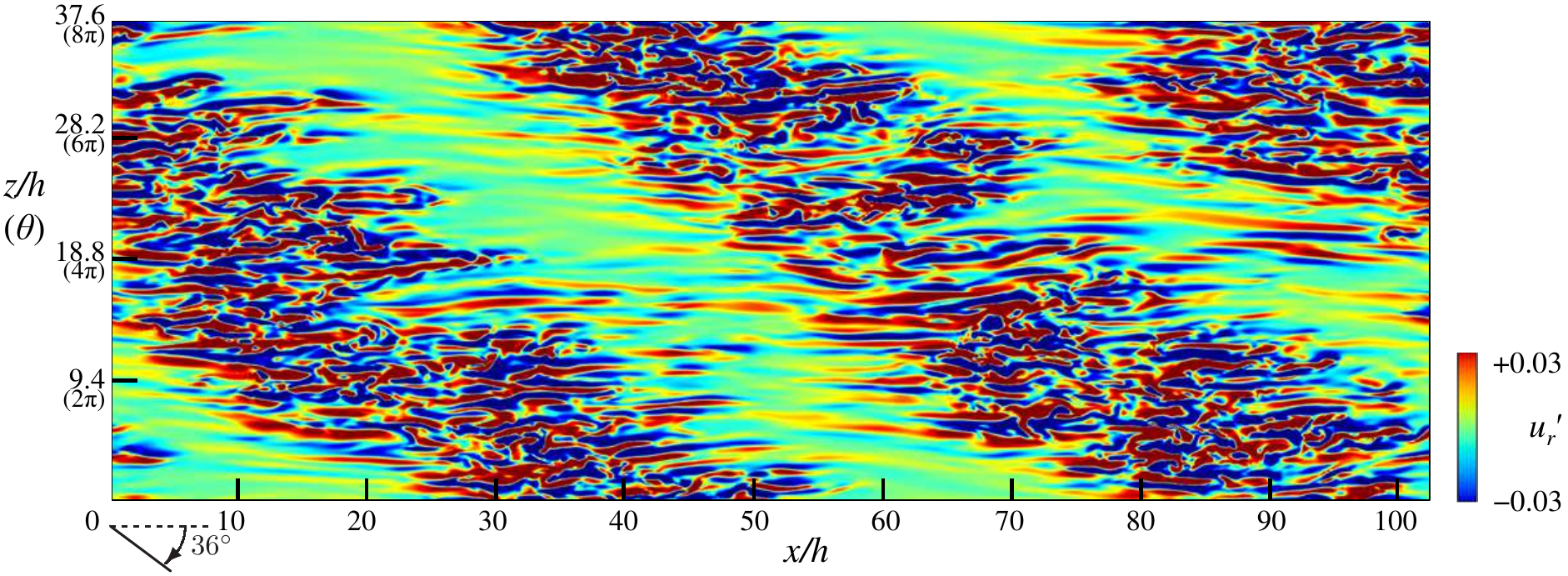} \\ \vspace{0.5em}
\end{center}
\caption{Simulations with artificial azimuthal extension $L_\theta = 8\pi$. Contours show instantaneous wall-normal velocity fluctuations $u_r^\prime$ in the $x$-$z$ (or $x$-$\theta$) plane at mid-gap for $\eta$ = 0.5. These flows are in a statistically steady state. (a) $\Rey = 387.5$, irregular laminar patches; in (b), $\Rey = 350$, helical turbulence.}
\label{fig:tran05_8pi}
\end{figure}

Figure~\ref{fig:tran05_8pi} shows two-dimensional distributions of the wall-normal velocity in the $x$-$z$ (or $x$-$\theta$) plane at mid-gap.
At $\Rey = 387.5$, laminar patches can be found, and there are no strong differences between the cases $L_\theta = 2\pi$ (figure~\ref{fig:tran05}b) and $L_\theta = 8\pi$ (figure~\ref{fig:tran05_8pi}a).
By reducing $\Rey$ further down to 350, helical turbulence occurred for $L_\theta = 8\pi$ (figure~\ref{fig:tran05_8pi}b), whereas the flow was fully laminar at the same value of  $\Rey$ for $L_\theta=2\pi$.
The pitch angle $\alpha_\theta$ is $36^{\circ}$ with respect to the streamwise direction, which appears comparable to the value of $29^\circ$ reported for $\eta = 0.8$. Such a slight difference is trivial in the present study, because $\alpha_\theta$ is dictated by the domain aspect ratio via the geometric relation $\alpha_\theta = \tan ^{-1} (pL_z/mL_x)$, with $p$ and $m$ integers.
This helical turbulence regime is found to relaminarise at $\Rey = 325$, which is similar to the case $\eta = 0.8$ and to pCf.

As a temporary conclusion, {unlike aPf,} the flow in realistic aCf does not feature any helical puff or stripe for $\eta=0.5$. Nevertheless, artificial azimuthal extension of the numerical domain, without change of curvature, reveals that helical stripes can in fact be sustained if the perimeter reaches $8\pi$. This suggests
that the lack of occurrence of helical large-scale structures in aCf with $\eta=0.5$ is not due to the wall curvature, but to geometric confinement.

\subsection{Low $\eta=0.1$}

\subsubsection{Nominal parameters}

\begin{figure}
\begin{center}
(a)\includegraphics[width=0.95\linewidth]{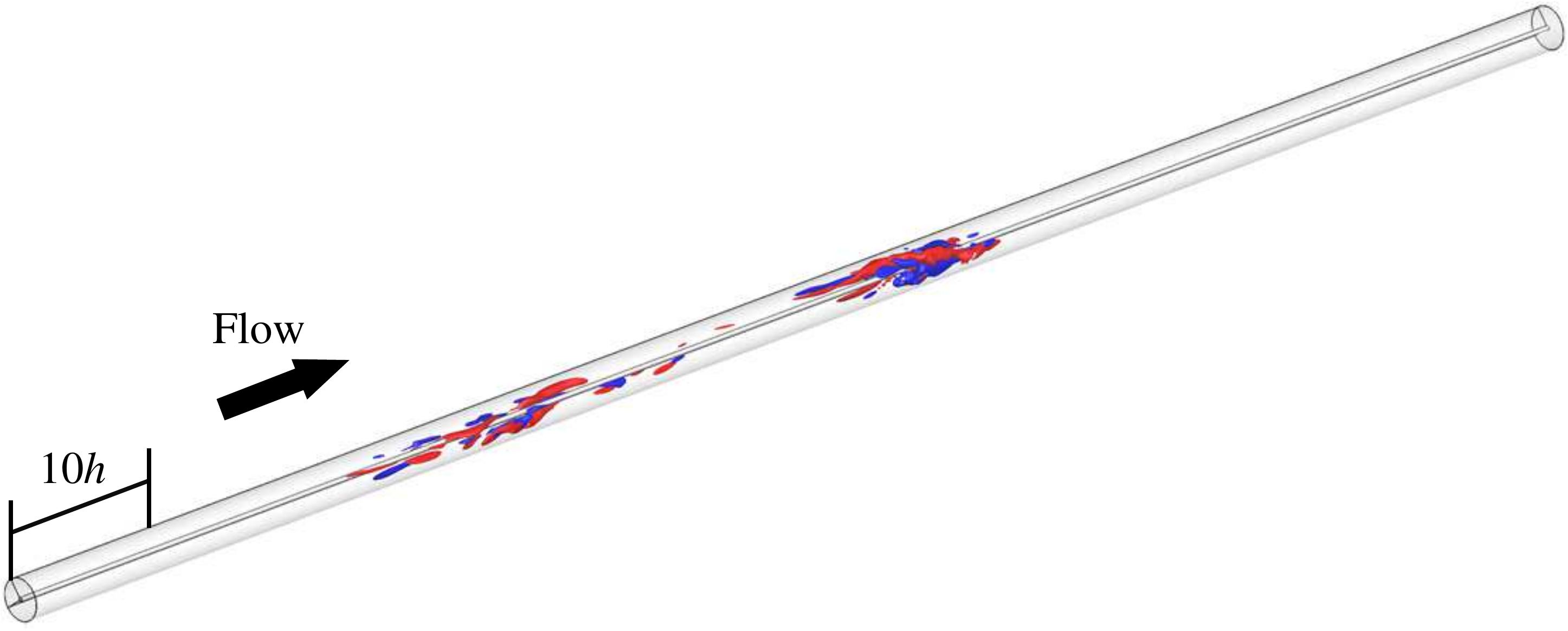} \\ \vspace{0.5em}
(b)\includegraphics[width=0.95\linewidth]{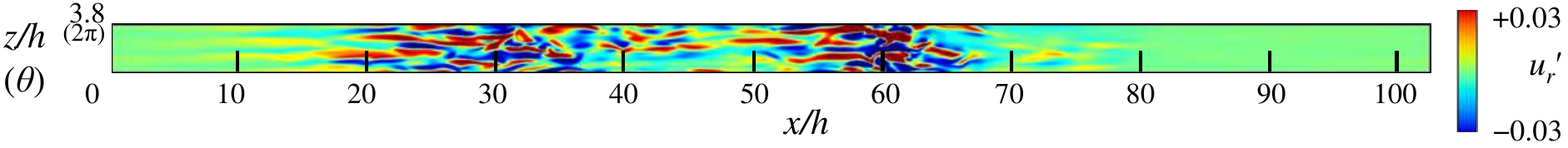} \\ \vspace{0.5em}
\end{center}
\caption{Same as figure~\ref{fig:tran08}, but for $\eta = 0.1$ and $\Rey = 400$, featuring puff-like structures. Only half the axial length of the computational domain is displayed.}
\label{fig:tran01}
\end{figure}

For $\eta = 0.1$, no oblique laminar-turbulent interface was observed in our simulations {either}, neither as a permanent regime nor even transiently. 
By direct analogy with aPf, for which no helical structure was ever detected below $\eta \le 0.3$, the immediate conclusion is that $\eta=0.1$ is too low for banded turbulence to occur.
There are, however, clear signs of laminar-turbulent coexistence {: c}oherent structures localised in the streamwise direction only, similar to puffs in pipe flow, have been visualised for $\Rey$ strictly below 400. These structures are referred to as ``puff-like'' because there are both common points and dissimilarities with the puffs found in pipe flow intermittency. 
Several such puff-like structures are shown in figure~\ref{fig:tran01} to coexist for $\Rey = 400$. Their long-time dynamics suggests that they are in equilibrium; however, {if these coherent structures decay after a sufficiently longer time, the flow goes back to the globally laminar state and puffs are only metastable objects}. At slightly higher $\Rey$, splitting of some of the puffs is also observed. 
All these features have been reported as robust statistical properties of cylindrical pipe flow \citep{Shimizu14}. 

\begin{figure}
\begin{center}
\includegraphics[width=0.95\linewidth]{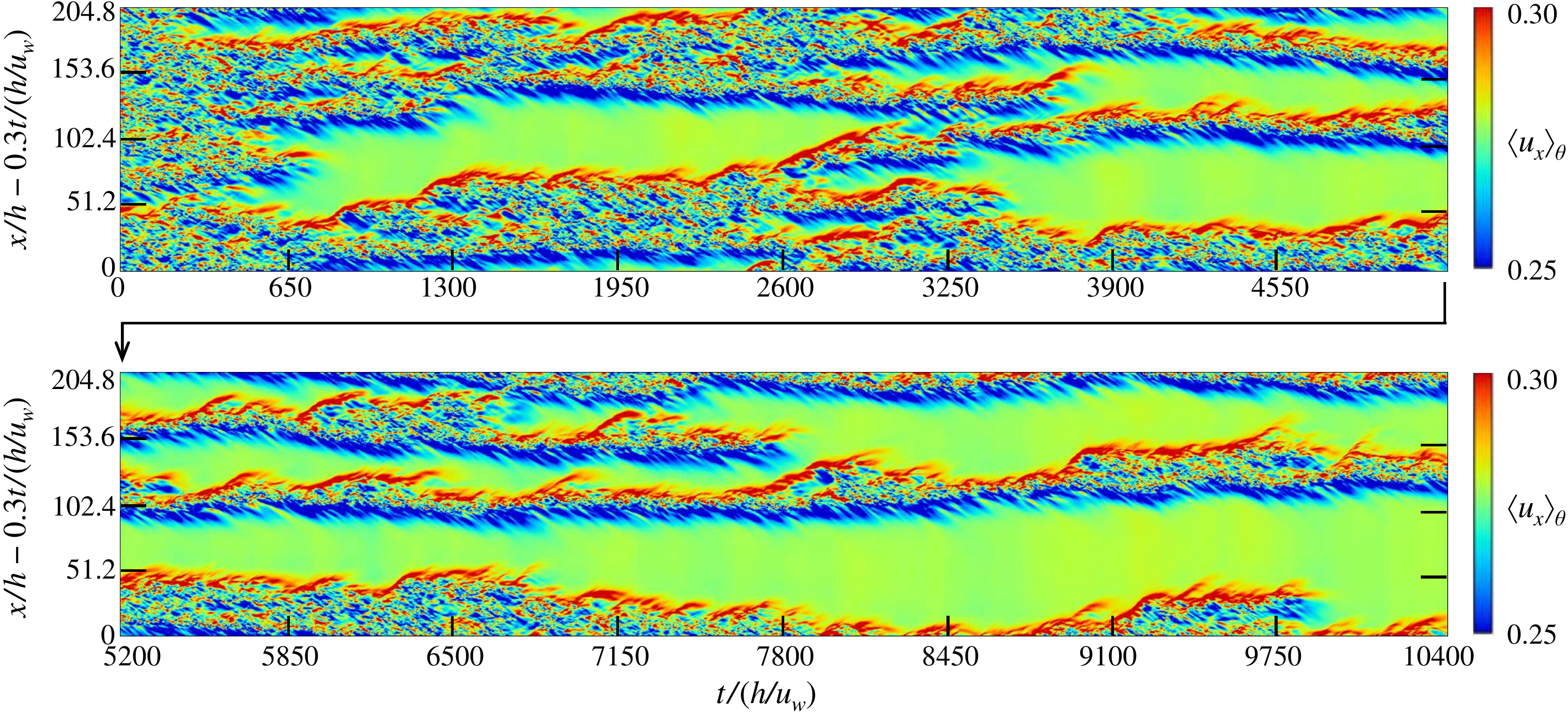}\\
(a) $\Rey = 400$ \vspace{1em}\\
\includegraphics[width=0.95\linewidth]{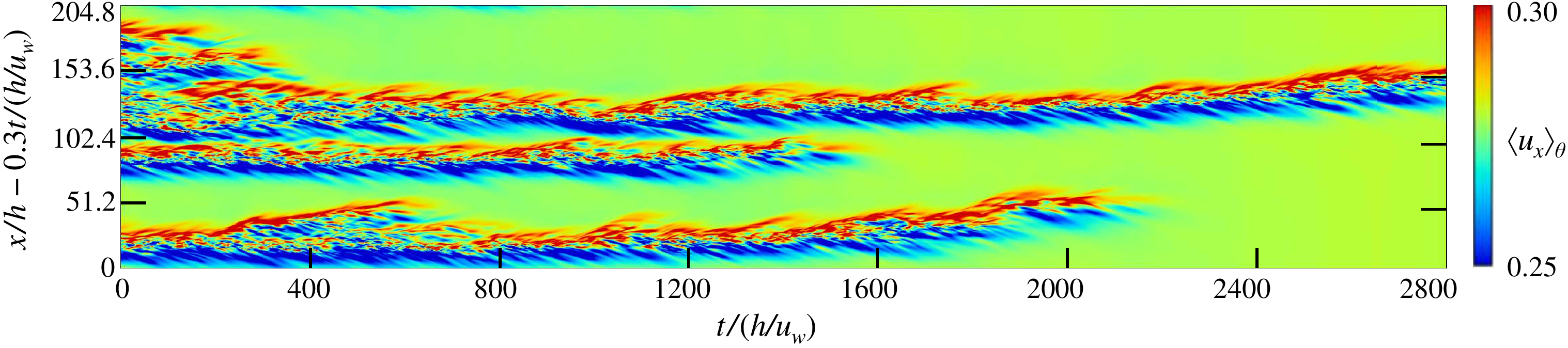}\\
(b) $\Rey = 395$ \vspace{0.5em}
\end{center}
\caption{Spatiotemporal diagrams of the azimuthally-averaged streamwise velocity $\langle u_x \rangle_\theta$ at mid-gap for $\eta = 0.1$ and $L_\theta = 2\pi$. The vertical axis shows the streamwise distance in a frame of reference moving at streamwise speed $0.3u_w$.}
\label{fig:std}
\end{figure}

Figure~\ref{fig:std} {illustrates} the spatio-temporal intermittency found for $\eta=0.1$. Two space-time diagrams for $\Rey=400$ (a) and 395 (b) show the azimuthally-averaged streamwise velocity $\langle u_x \rangle_\theta=\int u_x {\rm d} \theta/2\pi$, evaluated at mid-gap as a function of {$x$ and $t$}. In both cases, perturbations appear close to the laminar-turbulent interface{. They} can propagate in both upstream and downstream directions from the core of the puff towards the laminar flow. This makes the puff appear more symmetric than its pipe flow counterpart, where most perturbations are advected towards decreasing pressure \citep{Shimizu14}.
For the case  $\Rey=400$ (figure~\ref{fig:std}a), the flow apparently settles to a statistically steady state { for $t \ge 5000h/u_w$}. One can identify locally {transient quasi-laminar regions}, but no clear wavelength or patterning emerges.
It should be pointed out that no quasi-1D shear flow, has ever shown convincing patterning properties with well-defined wavelengths.

For $\Rey = 395$, all turbulent puff-like structures {are found to} decay.
According to the measurements above, the critical point for $\eta = 0.1$ lies between $\Rey = 400$ and 395.
If normalised by the hydraulic diameter ($2h$) and the ``centreline velocity'' $u_w$ as in pipe flow, this critical point corresponds to $4Re=1590\pm10$  {lower than the value of $2040 \pm 10$ for cylindrical pipe flow} \citep{Avila11}.

\subsubsection{Azimuthally extended system}
\label{sec:16pi}

\begin{figure}
\begin{center}
(a)\includegraphics[width=0.95\linewidth]{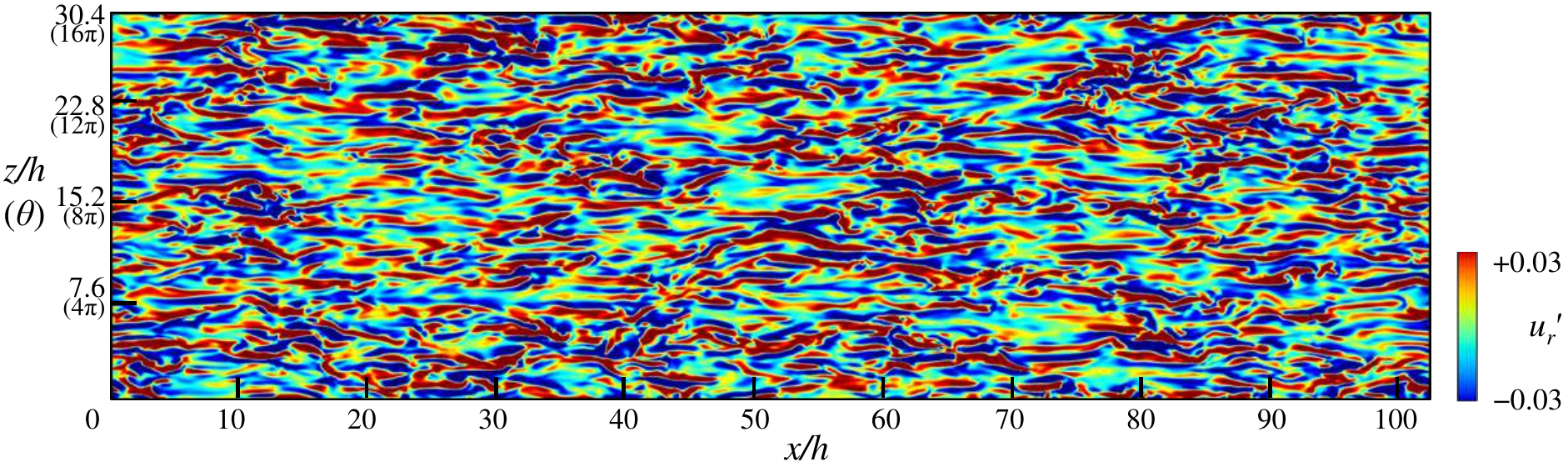} \\ \vspace{0.5em}
(b)\includegraphics[width=0.95\linewidth]{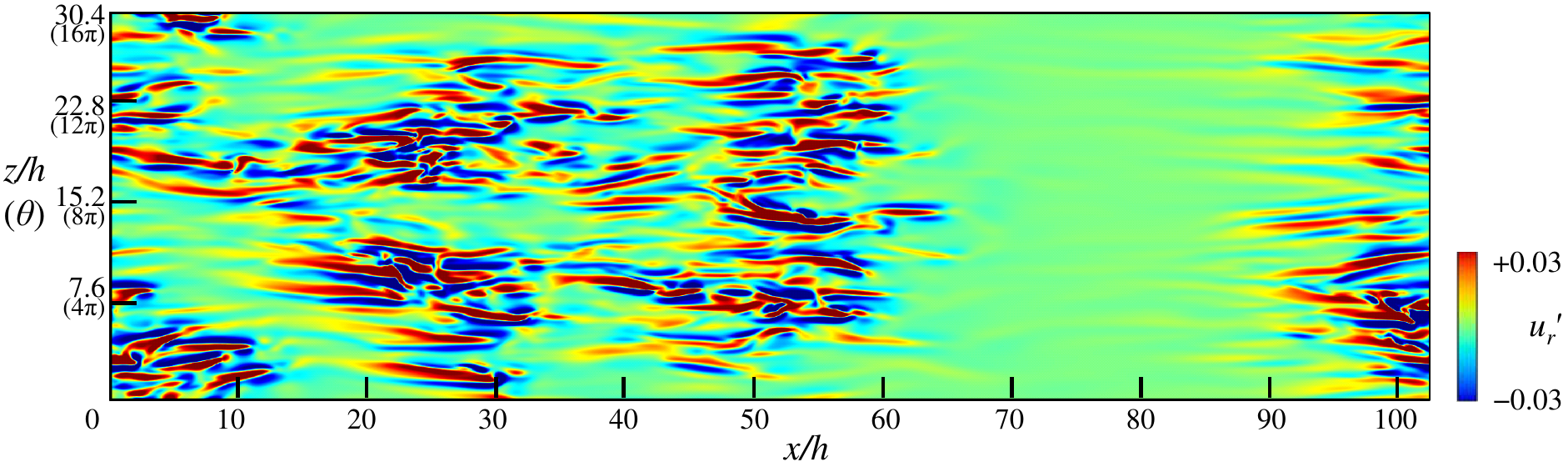} \\ \vspace{0.5em}
\end{center}
\caption{Instantaneous visualisation of wall-normal velocity fluctuations $u_r^\prime$ at mid-gap for $\eta$ = 0.1, $L_\theta = 16\pi$, and at (a) $\Rey = 400$ and (b) $\Rey = 275$. {The supplementary movie shows the time evolution during an arbitrary $800h/u_w$ of the flow visualised in (b), demonstrating a new regime of dynamic laminar-turbulent coexistence.} }
\label{fig:tran01_16pi}
\end{figure}

As {before}, we wish to go beyond the mere observation that no oblique pattern forms for $\eta=0.1$ and explain which mechanism, confinement or curvature, is responsible for the different structuration of the turbulent flow at its onset. We hence report direct numerical simulations for $\eta = 0.1$ in an azimuthally extended numerical domain with $L_\theta = 16\pi$ ($n=8$ times longer than the nominal circumference). In this case, $(L_{zi}, L_{zc}, L_{zo}) = (5.58h, 30.4h, 55.8h)$ and the resolution is kept to $N_\theta = 512$ {with $\Delta z^+ \le 3$}. {We note that the concept of `extended system' depends here highly on the radial position. This is an unavoidable geometric consequence of the relative smallness of the inner rod.} One effect of the azimuthal extension is again a marked shift of {$Re_g$}: the flow {does not return} to laminar even at $\Rey = 275$, a value much lower than $Re_g$ for $L_\theta = 2\pi$. Since the numerical resolution is still satisfying according to table \ref{tab:condition}, the shift in $Re_g$ is not attributed to the coarser resolution but to the change in boundary conditions. The nominal aCf with $\eta = 0.1$ relaminarises at $\Rey \approx 400$, this onset Reynolds number is strongly lowered for larger $L_\theta = 16\pi$ (figure \ref{fig:tran01_16pi}a) since relaminarisation in finite time has been observed for all $\Rey \le 262.5$. {For the} values of $Re$ where turbulence was detected, transiently or sustained, the flow displays laminar-turbulent coexistence, but never displays any helically-shaped turbulence, as can be seen in figure \ref{fig:tran01_16pi}b (see also the supplementary movie available at https://doi.org/10.1017/jfm.2019.666, which highlights the dynamical behaviour of the new flow regime).
Note that the {average streak width remains  unaffected} by the azimuthal extension beyond 2$\pi$. Given the very large perimeter of the outer wall in units of $h$, large enough in principle to accommodate one wavelength of a turbulent band, the absence of helical band, or equivalently of azimuthal large-scale flow, cannot be assigned to the azimuthal confinement, even though the inner perimeter is itself too small to accommodate this wavelength. We thus deduce that, unlike for $\eta=0.5$, the wall curvature is responsible for the inability of the geometry to sustain helical stripe patterns for $\eta=0.1$. 

\begin{figure}
\begin{center}
(a)\includegraphics[width=0.85\linewidth]{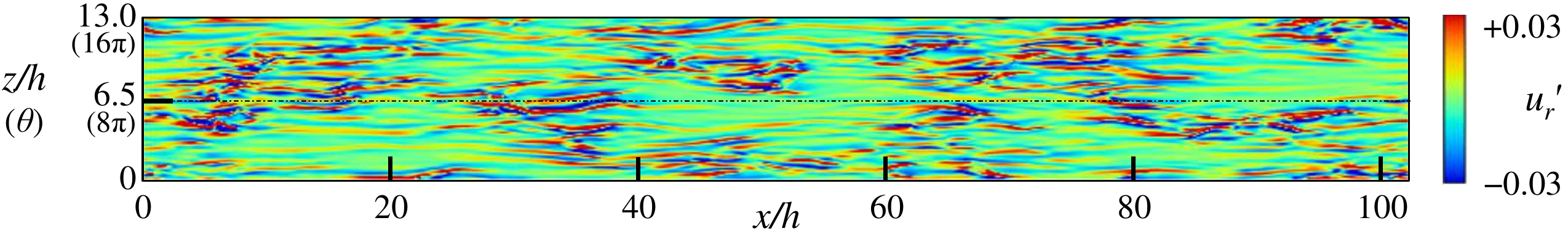} \\ \vspace{0.5em}
(b)\includegraphics[width=0.85\linewidth]{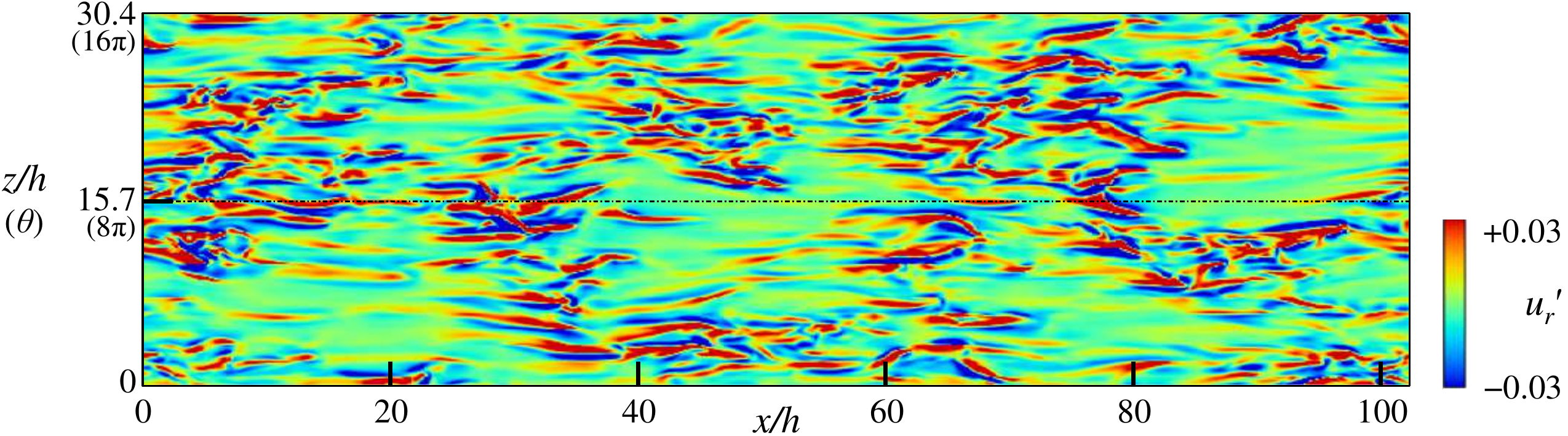} \\ \vspace{0.5em}
(c)\includegraphics[width=0.85\linewidth]{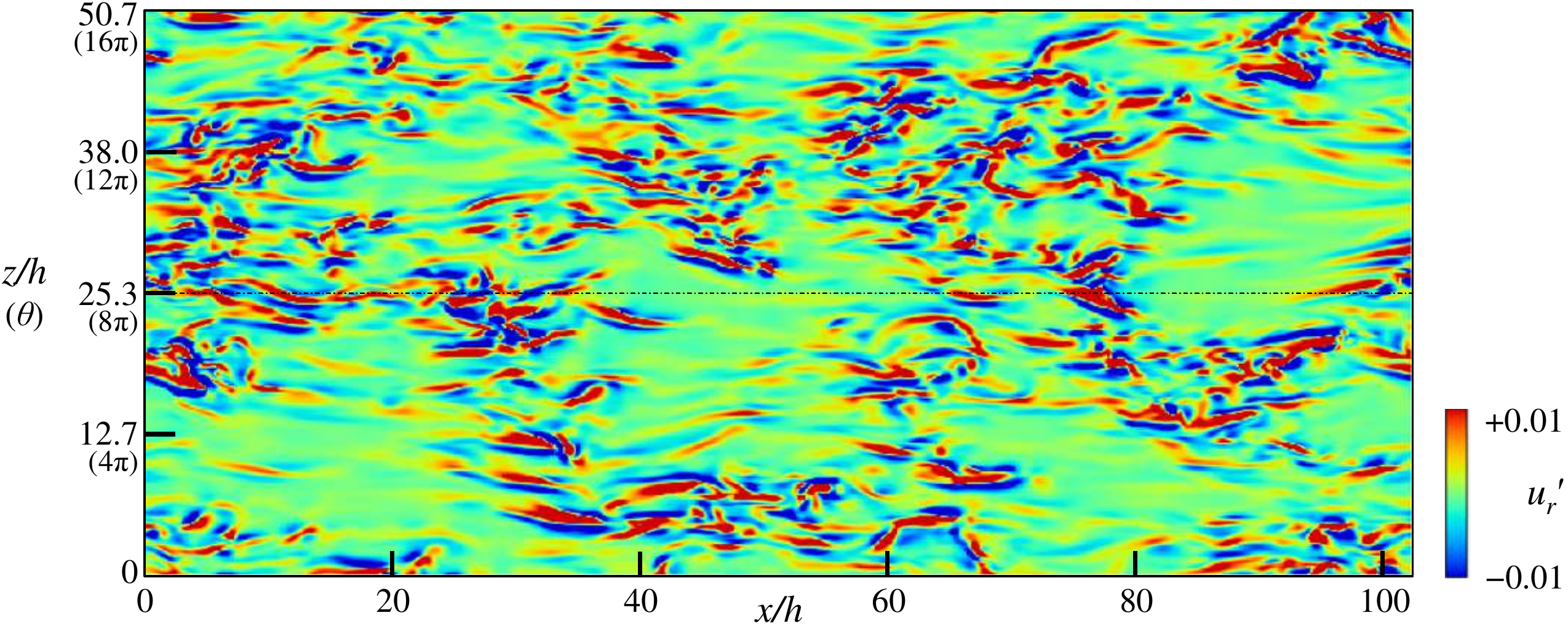} \\ \vspace{0.5em}
\end{center}
\caption{{Instantaneous visualisation of wall-normal velocity fluctuations $u_r^\prime$ in $x$-$\theta$ planes at three different radial positions for $\eta=0.1$, $L_\theta=16\pi$, and $Re =300$. From top to bottom : (a) $y \approx 0.15$ (near the inner wall) , (b)  $y \approx 0.5$ (mid-gap), and (c) $y \approx 0.9$ (near the outer wall).}}
\label{fig:r-dependence}
\end{figure}

Figure~\ref{fig:r-dependence} shows visualisations of the same {turbulent flow for $\eta=0.1$} at three different radii. It confirms that the topology of the flow is only weakly dependent on the radial value. 
Although azimuthal circumferences ($L_z/h$) differ between the three panels, the observed irregular patterns of localised turbulence are very similar to each other and match well when plotted {versus $\theta$}. 
As also confirmed in the supplementary movie, small-scale structures in turbulent patches seem to penetrate the entire gap from the inner to the outer wall. The intensity of velocity fluctuations in the region near the outer wall is weaker than near the inner wall, as also shown in figure~\ref{fig:r-dependence} (using different colourbars). The figure also makes the absence of large-scale helical pattern {clear} at any radial position. There is always a possibility that for sufficiently large azimuthal extension, both inner and outer perimeters become large enough to accommodate long-wavelength turbulent patterns again, yet with a very small pitch angle that can not be captured for smaller $L_{\theta}$.  However, given the present low value of $\eta$, this would imply {even longer computational domains and hence very expensive simulation}. This hypothesis is hence not considered in what follows.

\begin{figure}
\begin{center}
\includegraphics[width=0.95\linewidth]{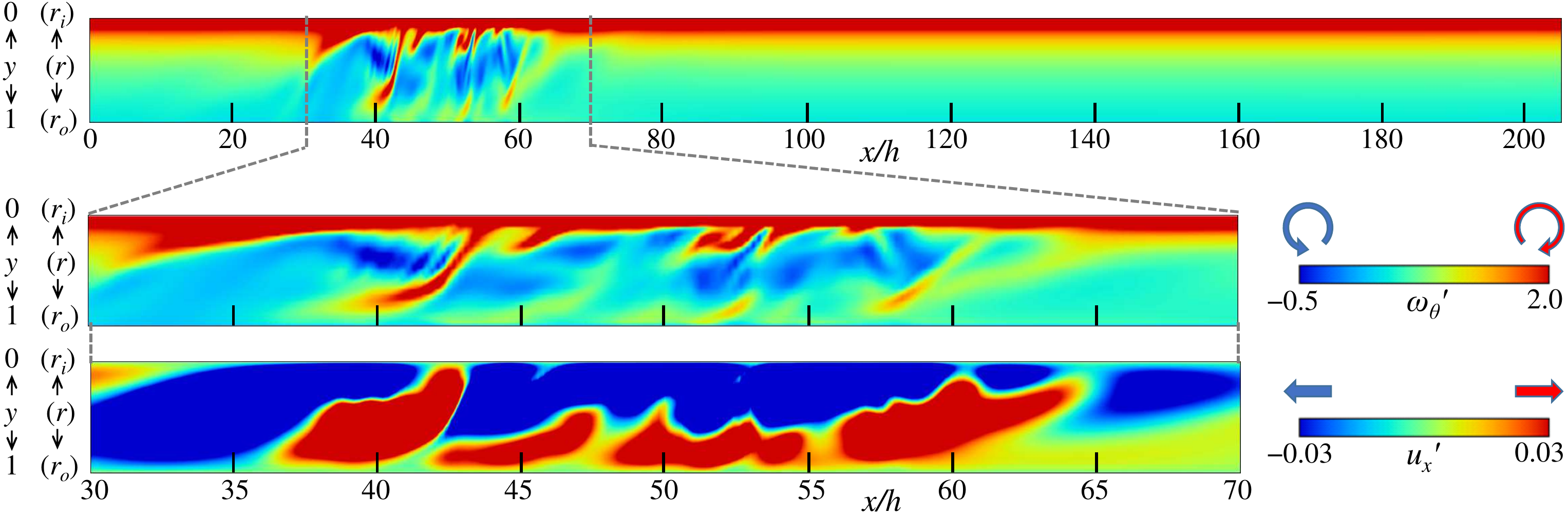}
\end{center}
\caption{Typical snapshots of azimuthal vorticity $\omega'_\theta$ (two top rows) and streamwise velocity fluctuation $u'_x$ (bottom row) distributions in an $x$-$r$ plane of aCf for $\eta=0.1$ and $\Rey = 395$ with nominal $L_\theta = 2\pi$. The inner rod (top boundary) moves towards the right of the figure.}
\label{fig:xr-plane}
\end{figure}

Among the many implications of wall curvature, an interesting one which can be checked for in simulations, in analogy with the corresponding laminar profiles, is the statistical asymmetry of the turbulent flow with respect to the mid-gap (this symmetry is exact only in the limit $\eta \rightarrow 1$). Figure~\ref{fig:xr-plane} shows the azimuthal vorticity and streamwise velocity inside an arbitrary meridian $x$-$r$ plane. The moving inner rod is located at the top of the figure while the outer rod lies at its bottom. {Strong mean gradients of azimuthal vorticity} (the ``boundary layer") are found near the inner wall only. This boundary layer triggers ejections of vorticity towards the outer wall, at various places corresponding to the location of the turbulent patch. This calls for a comparison with the regeneration mechanisms discussed in the context of pipe flow puffs \citep{Shimizu09,Duguet10b,Hof10}, where transverse vorticity acts as a source of perturbations that re-energises the turbulent puff against its own viscous decay. In these studies, the perturbations triggered by the instability of the corresponding shear layer are advected downstream, out of the puff. In the present context of aCf for $\eta=0.1$, the perturbations emanating from the shear layer at the inner rod are advected upstream in the frame moving with the inner wall. This suggests a turbulence regeneration process of a different kind. More work is needed to relate rigorously this observation to the short coherence length of these patches, as will be discussed in the next section. Section~\ref{sec:statistics} is devoted to a deeper quantitative analysis of the statistical properties of all the laminar-turbulent coexistence regimes identified in this section.

\section{Statistics}
\label{sec:statistics}

The statistics presented in this section have been gathered from the numerical runs described in the previous section. The cases simulated
with $L_{\theta}=2\pi$ are summed up in table \ref{tab:condition}, while the two azimuthally extended simulations, $\eta=0.5$ with $L_{\theta}=8\pi$ and
$\eta=0.1$ with $L_{\theta}=16\pi$ are described in \S~\ref{sec:8pi} and \ref{sec:16pi}, respectively. Importantly, all spatial statistics presented here are averaged over the
whole spatial domain (or sometimes over the mid-gap only), but are not conditional : these are not statistics over the turbulent zones only, rather global statistics
of laminar-turbulent coexistence seen as a single regime. 

\subsection{Mean velocity profile}

\begin{figure}
\begin{center}
\includegraphics[height=5.5cm]{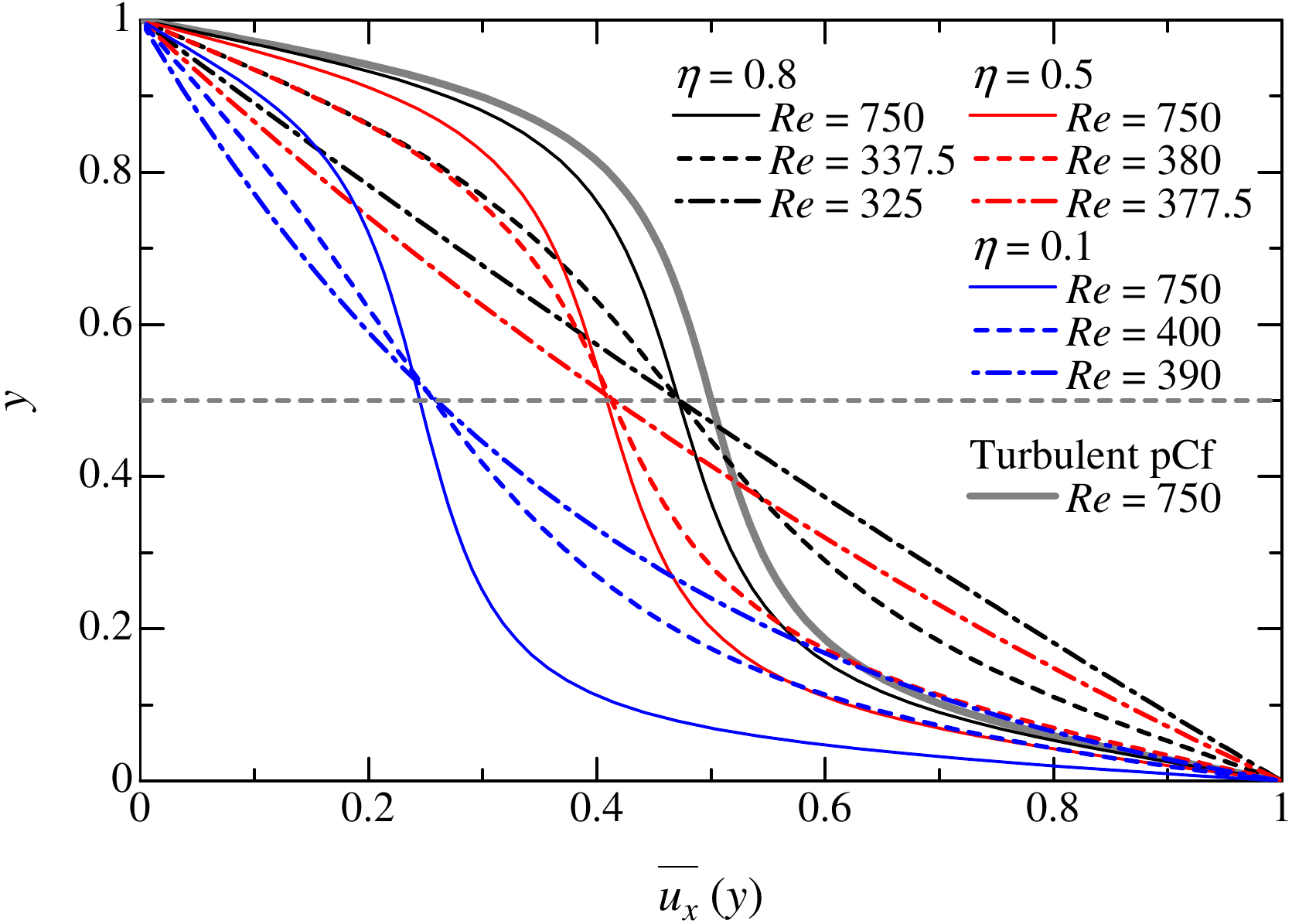}
\includegraphics[height=5.5cm]{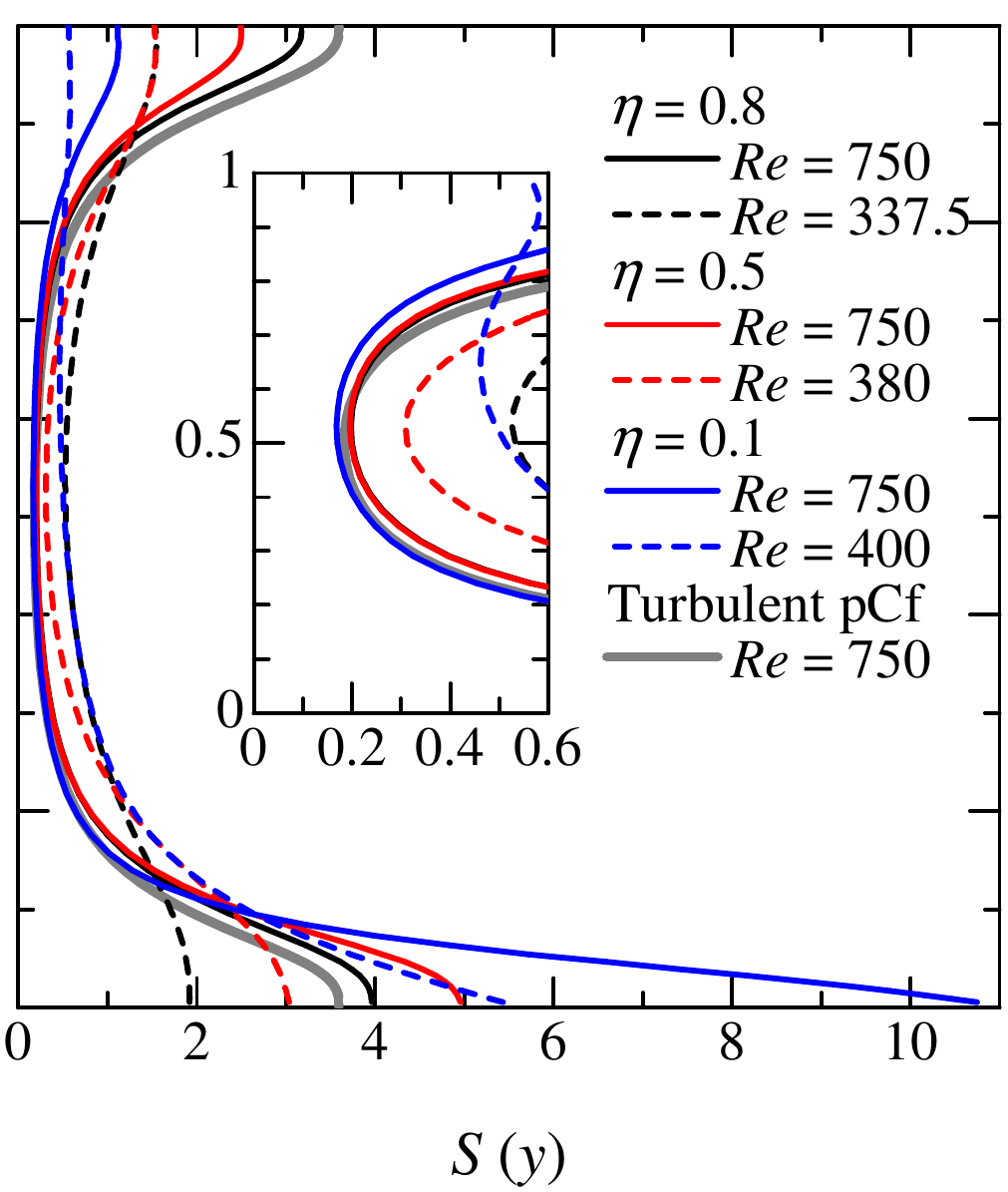}
\end{center}
\caption{Mean profiles of the streamwise velocity (left) and its velocity gradient $S$ (right) for various $\eta$ and $Re$. Black, red, and blue lines show $\eta =$ 0.8, 0.5, and 0.1, respectively. The dashed-dotted lines are of laminar and in agreement with the theoretical solution. For reference, a velocity profile of {turbulent} pCf \citep{Tsukahara06} {is also shown}.}
\label{fig:uvelo}
\end{figure}

The mean streamwise velocity profiles are displayed in figure~\ref{fig:uvelo}.
Here, the vertical axis $y$ ($= r - r_i$) is the radial distance to the wall measured from the inner cylinder.
The overbar denotes time-averaging in space over the two variables $x$ and $z$, and in time over a time horizon $T$.
Let $Re_g$ be the lowest value of $Re$ where turbulence is found for each value $\eta$, and let us assume by convention that it indicates relaminarisation for $\Rey<Re_g$, while in the other cases turbulence could sustain at least until time $T$, with $T$ limited here to $O(10^4)$. Our results indicate that $Re_g$ has a strong dependence on $\eta$. Mean velocity profiles, evaluated here only for $\Rey$ close to $Re_g(\eta)$, are {also} affected by $\eta$. The dashed-dotted lines indicate the laminar profiles, in agreement with the theoretical solutions given in equation~(\ref{eq:base}) {as well as in figure~\ref{fig:ubase}}. We note that for $\eta=0.1$ for instance, the profile for $Re = 390$ deviates from the cluster formed by the other profiles reported for $400 \le Re \le 750$.
The last observation results from the inclusion \emph{de facto} of the laminar parts in the spatial averaging : any variation in the mean turbulent fraction (which will be documented in \S~\ref{sec:Ft}) impacts the mean velocity profile.

A clear trend emerges from all the non-laminar cases reported in figure~\ref{fig:uvelo} : the mean velocity profiles are S-shaped and all display an inflection point located near $y=0.5$. The symmetric property, which the limiting case of pCf also possesses, apparently persists in all aCf configurations despite the broken symmetry of the system with respect to the mid-gap. This does not {necessarily} imply shear instabilities yet. Recalling the novel kind of laminar-turbulent coexistence identified for low $\eta=0.1$, it is interesting to compare how the mean profile {evolves} with decreasing $\eta$. Other features of interest can be extracted from figure~\ref{fig:uvelo}, such as the mean shear {profile}, given by $S(y):=-\partial \overline{u_x}/\partial y$. 
As $\eta$ decreases, this velocity gradient becomes steeper at the inner cylinder than at the outer one. The profile also becomes more and more asymmetric with respect to the mid-gap $y=0.5$. A quantitative comparison is given for $\Rey=750$, a case for which statistics are not affected by laminar-turbulent coexistence. While $S(y)$ is approximately 0.2 at mid-gap for all values of $\eta$, its value ranges from 1 to 3 at the outer wall (decreasing with decreasing $\eta$) and from 4 to 11 at the inner wall (now increasing  with decreasing $\eta$). The trend is hence such that, as $\eta$ decreases towards 0.1, the mean profile gets more asymmetric, and resembles more and more a simple
linear boundary layer profile located near the inner rod, while the gradients at the outer rod are increasingly weak.

\subsection{Friction factor}

\begin{figure}
\begin{center}
\includegraphics[width=0.8\linewidth]{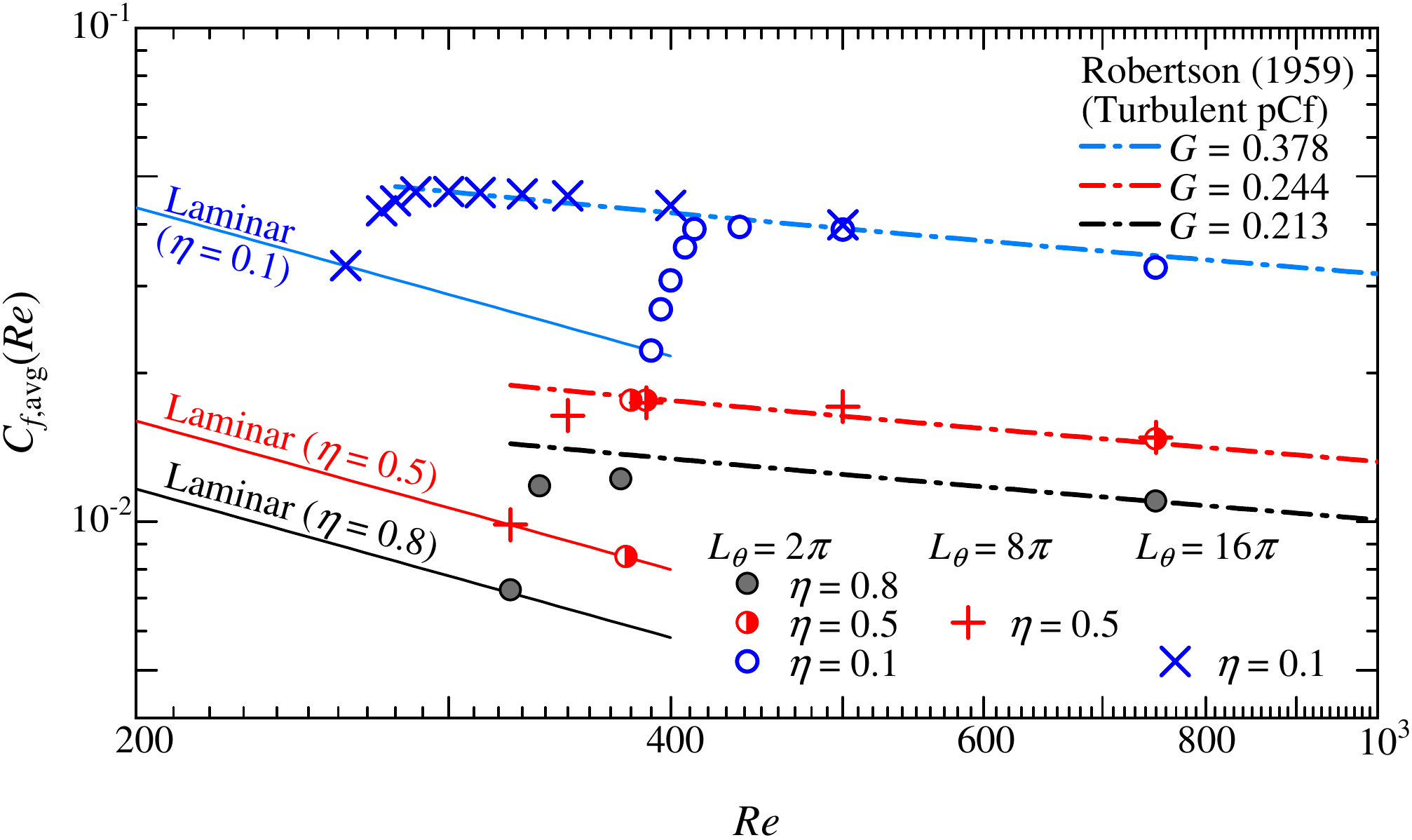}
\end{center}
\caption{Friction coefficient as a function of $\Rey$ including the cases of artificially extended $L_\theta$. Solid and dashed-dotted lines show equations~(\ref{eq:friction_laminar}) and (\ref{eq:robertson}), respectively.}
\label{fig:friction_plus}
\end{figure}

We present {here} calculations of the friction coefficient $C_f$ versus the {$Re$}, displayed in 
figure~\ref{fig:friction_plus}. To the authors' knowledge, no experimental study {has} examined {$C_f$} of aCf except {for} \cite{Shands80}, who {reported}  measurements in a wide range of Reynolds number including laminar and fully-developed turbulent regimes for low $\eta < 0.03$. $C_f$ for the present study is defined as :
\begin{equation}
 C_f = \frac{2\tau_w}{(\xi u_{w})^2},
 \label{eq:friction}
\end{equation}
where the coefficient $\xi$ stands for the ratio between the dimensional bulk velocity and the dimensional wall velocity $u_w$. Note that $\tau_w$ and $\xi$ are both defined as temporal averages. The introduction of the ratio $\xi$ is intended to capture the dependence of the bulk-mean velocity on $\eta$. For instance, $\xi$ should be 0.5 for the case $\eta=1$.
In this series of numerical experiments, we obtained $\xi$ = 0.46 ($\eta= 0.8$), $\xi$ = 0.39 ($\eta=0.5$), and $\xi$ = 0.20 ($\eta= 0.1$).
In figure~\ref{fig:friction_plus}, the plotted quantity $C_{f,\rm avg}$ is a weighted average between the inner and outer values ${C_f}$, {i.e.,}
\begin{equation}
 C_{f,\mathrm{avg}} = \frac{\eta C_{f,\mathrm{inner}} + C_{f,\mathrm{outer}}}{\eta+1},
 \label{eq:cf_avg}
\end{equation}
where each $C_f$ is computed {from} the wall shear stress $\tau_w$  {measured  locally}. The solid line shows the laminar law {given} by
  \begin{equation}
   C_{f,\mathrm{laminar}} = \frac{4}{\xi^2} \frac{1-\eta}{Re (1+\eta) |\ln \eta|},
   \label{eq:friction_laminar}
  \end{equation}
and the dashed-dotted line is the empirical law suggested by \cite{Robertson59} for turbulent pCf :
  \begin{equation}
   \sqrt{\frac{C_f}{2}} = \frac{G}{\log Re}.
   \label{eq:robertson}
  \end{equation}
Circles and crosses label the results with nominal and extended $L_\theta$, respectively.
{The constant $G$ is determined by a logarithmic fit of $\sqrt{C_f}$ directly from the simulations.}\\

We start by describing the results obtained for standard numerical domains with $L_{\theta}=2\pi$. We first note {that Robertson's formula}, borrowed from the {plane} Couette case, captures surprisingly well the turbulent {regime} for all values of $\eta$ as long as $Re$ is large enough (above 500).
A `transitional' region, i.e. a range comprising both non-laminar and non-fully-turbulent regions, emerges for $325 <\Rey< 400$ (for $\eta = 0.8$) and $395 < \Rey < 412$ (for $\eta = 0.1$).
As for $\eta = 0.5$, the turbulent flow becomes laminar at $\Rey = 377.5$ and, hence, no intermediate value of $C_f$ value between laminar and turbulent values was identified.
From figure~\ref{fig:friction_plus}, $Re_g$ can be approximated by 325, 380, and 400 for $\eta = 0.8$, 0.5, and 0.1, respectively.

In the cases of artificially extended $L_\theta$ (crosses in figure~\ref{fig:friction_plus}), the variations of $C_f$ exhibit different but still interesting trends. For $\eta = 0.5$, $Re_g$ of $L_\theta = 8\pi$ (red crosses) is lower than that of $L_\theta = 2\pi$ and the new value of $Re_g = 325$ is {similar} to the threshold Reynolds number reported in pCf by \cite{Duguet10}. For $\eta = 0.1$, the lower value of $Re_g$ has already been reported in the previous section. It is surprising that $C_f$ is very close to the fully turbulent value (see the blue crosses), even at $\Rey = 300$, although the {flow field is much more intermittent} than at higher $\Rey$ (as will be quantified by the turbulent fraction in \S~\ref{sec:Ft}).

\subsection{Two-point correlations}
\label{sec:tpc}

\begin{figure}
\begin{center}
\begin{tabular}{cc}
\begin{minipage}{0.32\textwidth}
\includegraphics[width=0.95\textwidth]{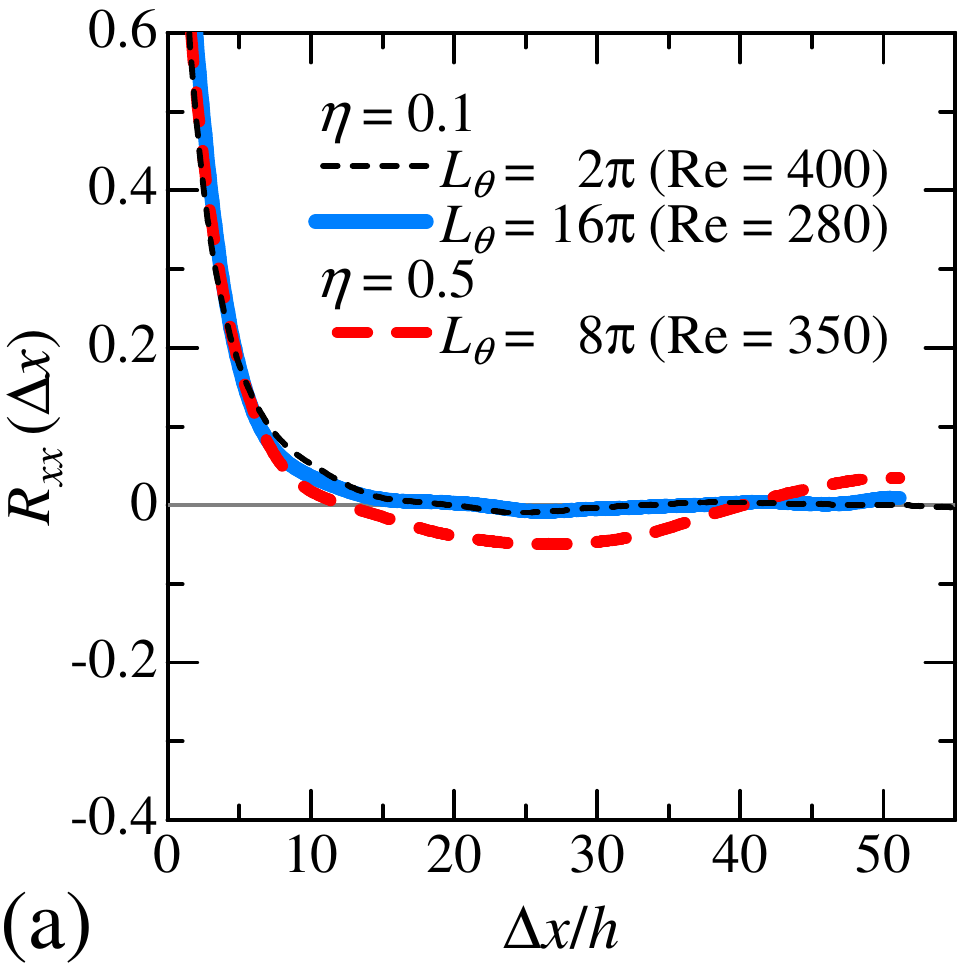}
\end{minipage} \,
\begin{minipage}{0.32\textwidth}
\includegraphics[width=0.95\textwidth]{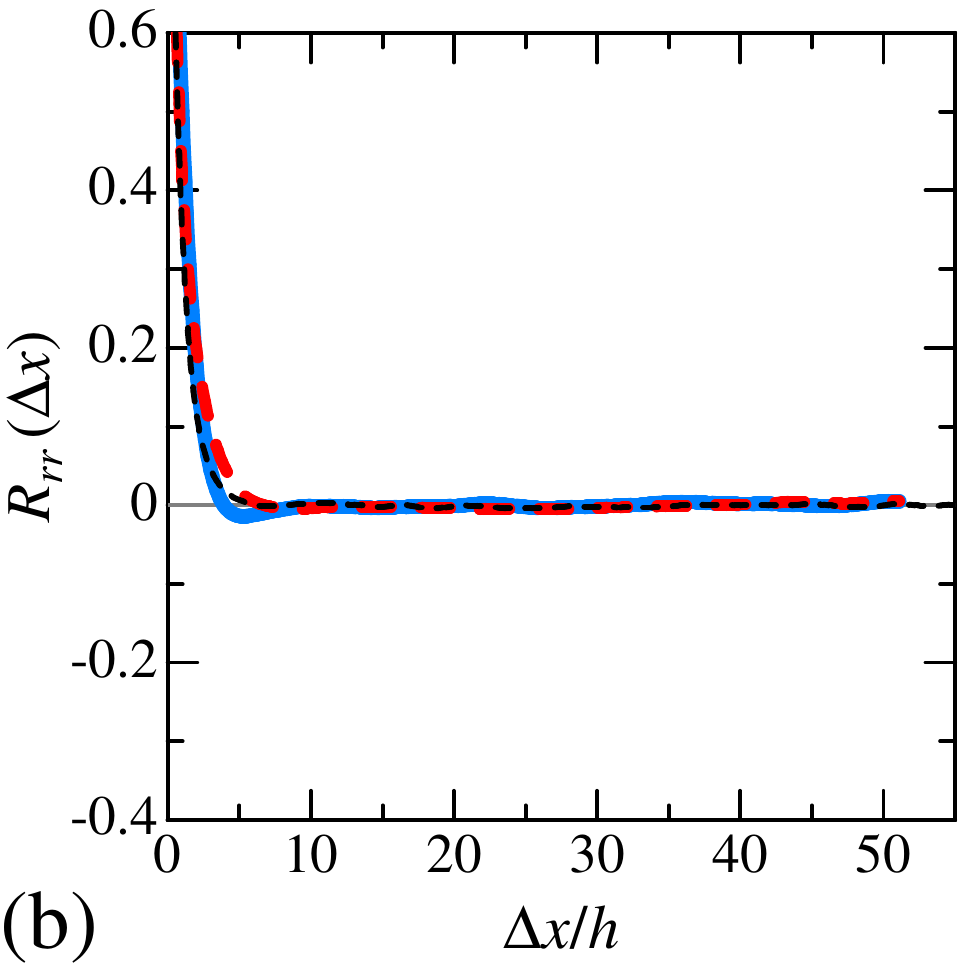}
\end{minipage} \,
\begin{minipage}{0.32\textwidth}
\includegraphics[width=0.95\textwidth]{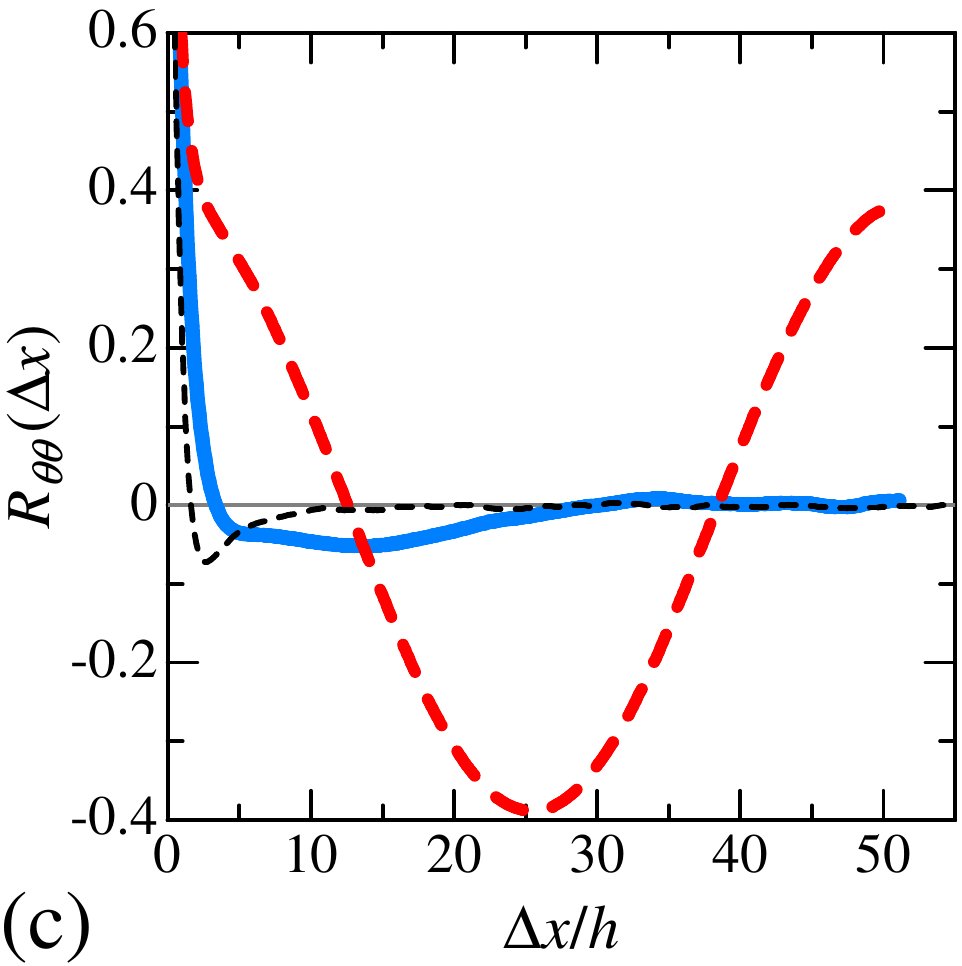}
\end{minipage} \vspace{1em} \\
\begin{minipage}{0.32\textwidth}
\includegraphics[width=0.95\textwidth]{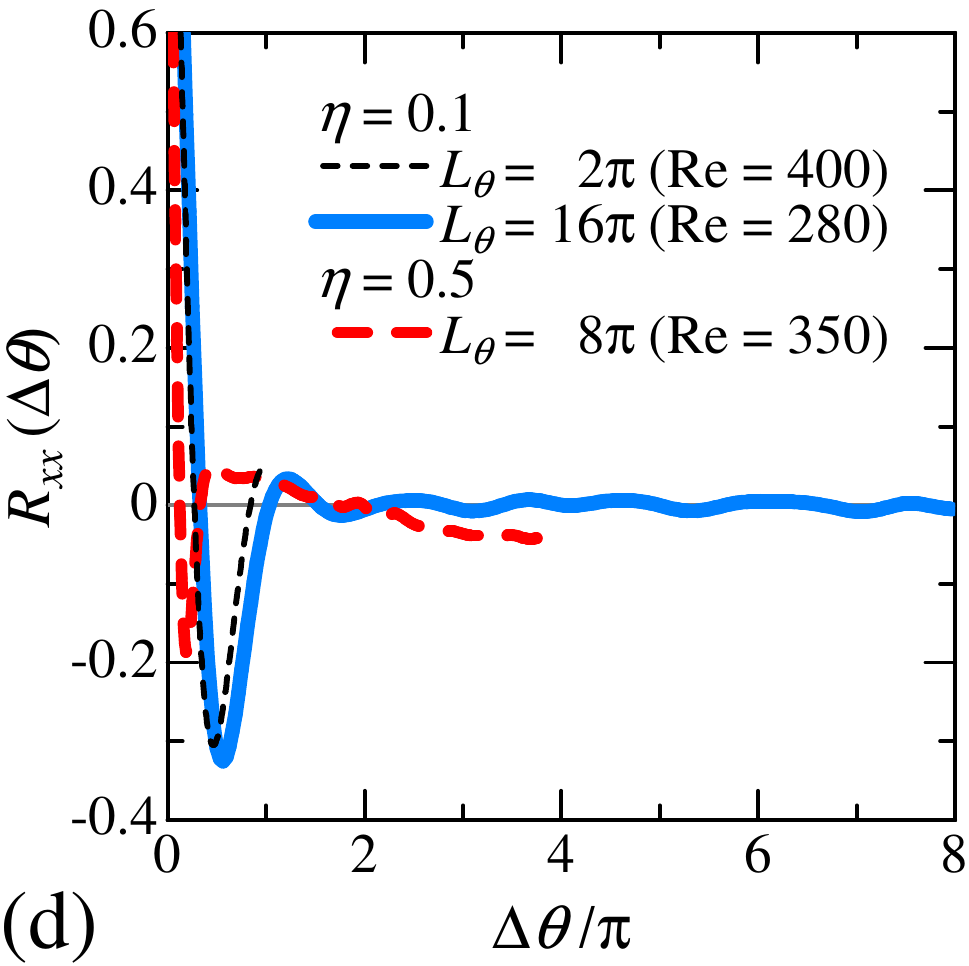}
\end{minipage} \,
\begin{minipage}{0.32\textwidth}
\includegraphics[width=0.95\textwidth]{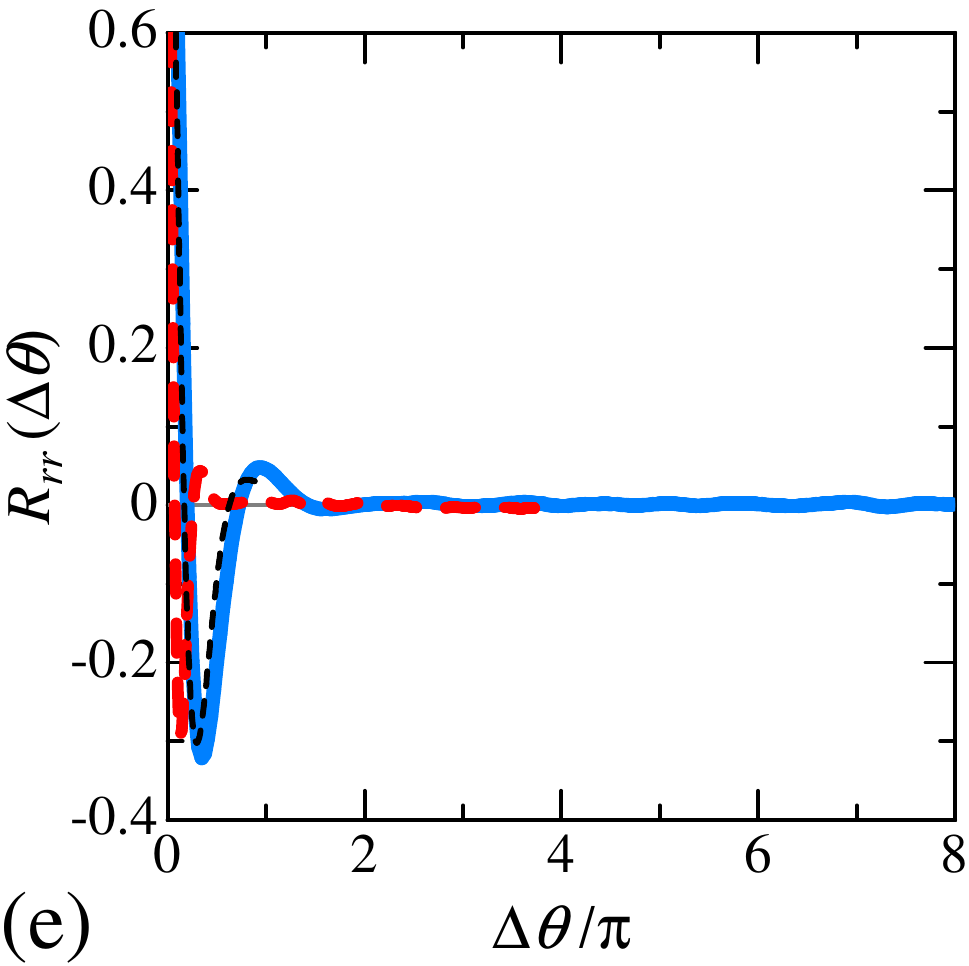}
\end{minipage} \,
\begin{minipage}{0.32\textwidth}
\includegraphics[width=0.95\textwidth]{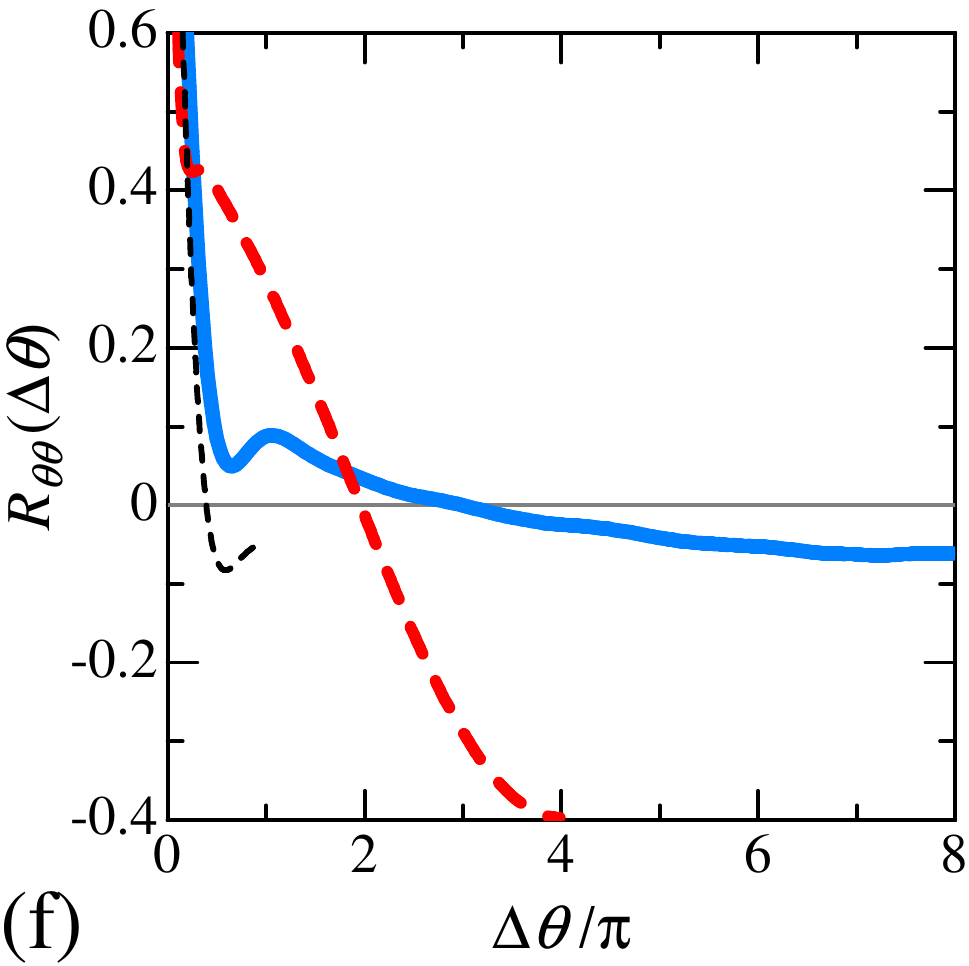}
\end{minipage}
\end{tabular}
\end{center}
\caption{Two-point correlation of velocity fluctuations as a function of either an axial lag $\Delta x$ (top row) determined by equation~(\ref{eq:xtpc}) [in (a--c)] or an azimuthal lag $\Delta \theta$ (bottom row) determined by equation~(\ref{eq:ztpc}) [in (d--f)] : $R_{xx}$ in (a, d), $R_{rr}$ in (b, e), and $R_{\theta\theta}$ in (c, f) represent autocorrelations of $u_x^\prime$, $u_r^\prime$, and $u_\theta^\prime$, respectively. The legend in (a) is also valid for (b--f).}
\label{fig:tpc}
\end{figure}

After describing statistics involving streamwise and azimuthal averages, we now describe the spatial correlations {in the case of laminar-turbulent coexistence (i.e. close to $Re_g$)}, restrained to the wall-parallel variables $x$ and $\theta$.
The {correlations of the velocity perturbations}, evaluated at a constant value of $y=y_{\rm ref}$ are classically defined, after normalisation, by : 
\begin{eqnarray}
R_{ii}(\Delta x) &=& \frac{\overline{u_i^\prime (x,y_{\rm ref},\theta) \cdot u_i^\prime (x+\Delta x,y_{\rm ref},\theta)}}{\overline{u_i^\prime (x,y_{\rm ref},\theta)^2}},
\label{eq:xtpc} \\
R_{ii}(\Delta \theta) &=& \frac{\overline{u_i^\prime (x,y_{\rm ref},\theta) \cdot u_i^\prime (x,y_{\rm ref},\theta+ \Delta \theta)}}{\overline{u_i^\prime (x,y_{\rm ref},\theta)^2}},
\label{eq:ztpc}
\end{eqnarray}
where the overbar $\overline{\left( \cdot \right)}$ again denotes averaging over $x$, $\theta$, and time. The value of $y_{\rm ref}$ is {approximately 0.5}, i.e. close to mid-gap. The present focus is on a neater characterisation of  the new regime identified for $\eta=0.1$ in the extended domain with $L_{\theta}=16\pi$. {To this end}, we compare in figure~\ref{fig:tpc} the autocorrelations of each velocity component as functions of the streamwise and azimuthal spatial lags $\Delta x$ and $\Delta \theta$, computed for $\eta=0.1$, to their {counterparts} for $L_{\theta}=2\pi$, as well as to the other extended case with $\eta=0.5$ and $L_{\theta}=8\pi$. {The results obtained for $L_{\theta}=2\pi$ are not significantly different between $\eta= 0.5$ and 0.1 (not shown here), because of the {common} absence of large-scale patterning apparent in figure~\ref{fig:tran05}}.

These correlations fall into two groups with well-defined trends in each group. The first trend {(panels a, b, d, e of figure~\ref{fig:tpc}) includes the autocorrelations of streamwise and radial velocity components for both streamwise and azimuthal lags : streamwise correlation curves differ weakly from case to case, but the axial correlation length is {less than 5$h$. This appears significantly shorter than the minimum puff spacing away from criticality, i.e. 20--30 diameters or larger values for lower Reynolds numbers \citep{Samanta11}.} The case with $\eta=0.5$ displays a comparable streamwise correlation length for the streamwise and radial velocity components, but shorter azimuthal correlations}, while for $\eta=0.1$ both values of $L_{\theta}$ yield similar lag angles of approximately $\pi$. 

The second trend {(panels c, f)} is for the correlations of the azimuthal velocity field : the correlation lengths in $x$ and $\theta$ for $\eta=0.5$  are very large, respectively $26h$ and $2\pi$, {because of the large-scale structure of the turbulent bands}.  For $\eta=0.1$ however, they shrink to respectively $\approx 5h$ and less than $\pi/2$. {For comparison with the case $\eta=0.1$, puffs in pipe flow in the lower transitional regime ($Re=2000$) display wider azimuthal correlations {and} longer streamwise correlations \citep{WillisKerswell2008}}.

These observations shed a new light on the regimes identified for $\eta=0.1$, regardless of the choice of $L_{\theta}$ : as far as the azimuthal component is concerned, the transitional regime for $\eta=0.1$ displays \emph{short-range} correlations not found for the higher values of $\eta$ investigated in this paper. The fact that the correlation decay is not modified between $L_{\theta}=2\pi$ and $16\pi$ suggests that this regime is robust and not an artefact of either confinement by boundary conditions or of the artificial numerical extension in $\theta$, {though confirmation for even larger $\eta$ would be welcome}. A rapid comparison with the correlations found in aPf \citep{Ishida17}, also close to the onset Reynolds number, confirms that they are all of the type encountered here for $\eta=0.5$ and 0.8. The regime documented in \S~\ref{sec:16pi} is hence, to the best our knowledge, a new unreported regime of laminar-turbulent coexistence with short-range interactions between the turbulent fluctuations. 

\subsection{Turbulent fraction}
\label{sec:Ft}

\begin{figure}
\begin{center}
\includegraphics[width=0.8\linewidth]{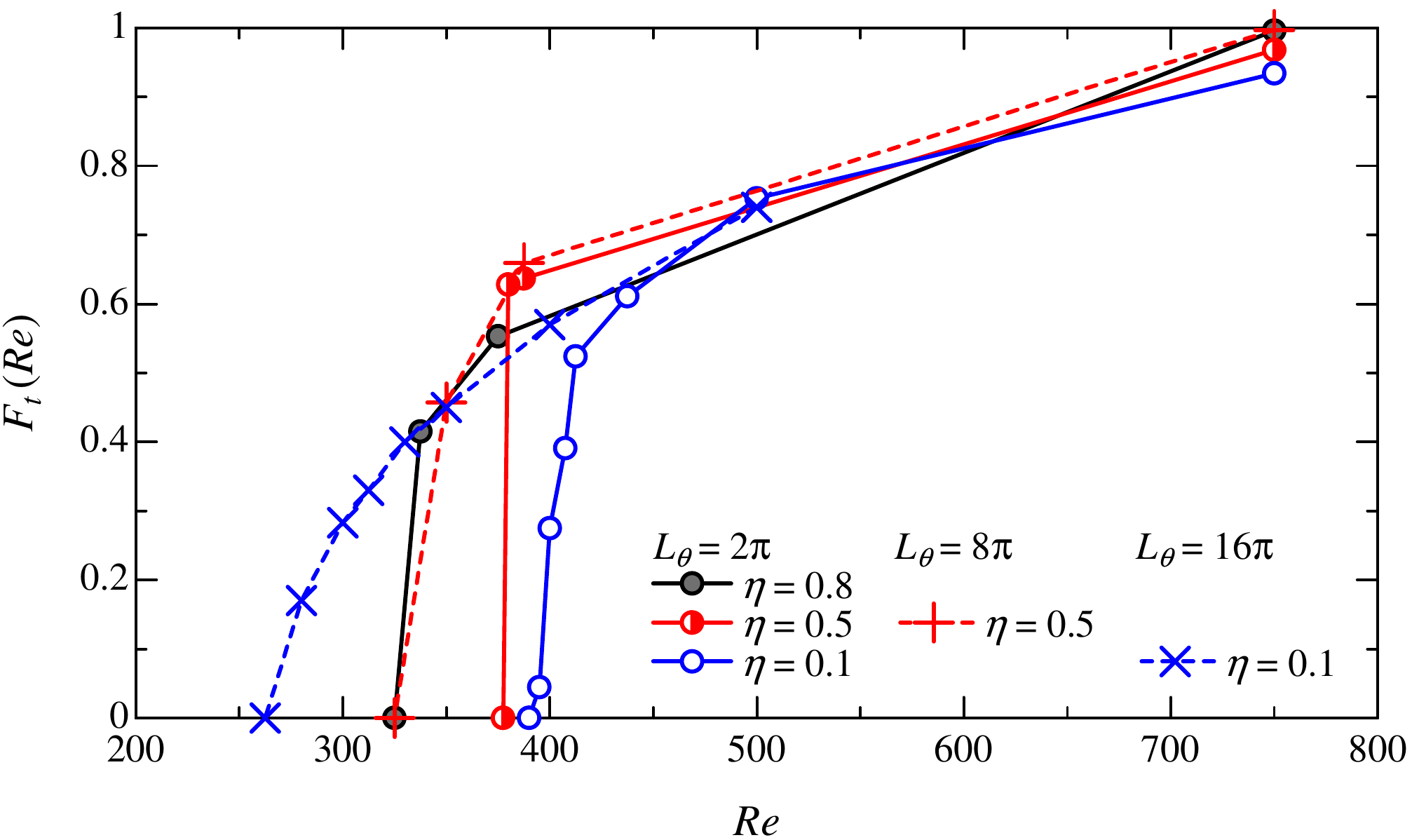}
\end{center}
\caption{Time-averaged turbulent fraction $F_t$ as a function of $\Rey$, nominal and artificially extended numerical domains with $L_\theta=2\pi,~8\pi$ and $16\pi$.}
\label{fig:ft_all}
\end{figure}

The turbulent fraction $F_t(t)$ measures the instantaneous amount of turbulence in the flow independently of its local intensity,
it is the natural way to quantify robust laminar-turbulent coexistence. If $F_t = 1$, the flow is turbulent everywhere, while $F_t = 0$ means that the flow is fully laminar. Figure~\ref{fig:ft_all} shows temporal averages of $F_t$, evaluated from $x$-$z$ plane data evaluated
at mid-gap. The local criterion for extracting laminar regions is based on thresholding the radial velocity : $|u_r^\prime| < 0.015$ indicates locally laminar flow, while $|u_r^\prime| > 0.015$ indicates locally turbulent flow.
As shown in figure~\ref{fig:ft_all}, for $Re=750$, $F_t > 0.9$ for all values of $\eta$, when the flow can be unambiguously described as fully turbulent.
$F_t$ monotonically decreases in average as $\Rey$ decreases, with a more significant decrease around $\Rey = 400$--500.
For $\eta = 0.8$ and 0.1, localised turbulent structures (helical turbulence and turbulent puff, respectively) occur and then $F_t$ approaches an average of 0.4. For $\eta = 0.5$, however, $F_t$ directly jumps from 0.6 to 0 due to the lack of localised turbulent structure.\\

We focus {then} on the azimuthally extended systems. For $\eta = 0.5$ $(L_\theta = 8\pi)$, helical turbulence occurs, but the slope of the curve $F_t(\Rey)$ and the pointwise values are directly comparable to the case $L_{\theta}=2\pi$ above the critical point. Only the few additional non-zero values of $F_t$ near the onset Reynolds number suggest a change in the curve of $F_t(Re)$, which can be linked to the release in azimuthal confinement. That change is compatible with continuous $F_t$ even at the origin, although the trend cannot be confirmed at that stage. The situation becomes clearer for $\eta=0.1$. For $\eta = 0.1$ $(L_\theta = 16\pi)$, $F_t$ shows values similar to the case $\eta = 0.1$ $(L_\theta = 2\pi)$ when $\Rey$ is {in the range} 400--750. However, below $Re = 400$, $F_t$ displays a much smoother decrease from values of 0.3--0.4 down to zero  for $Re \approx275$ (the last non-zero point has in fact $F_t \approx 0.15$). This is where helical turbulence appears for  $\eta = 0.8$ and $\eta = 0.5$ $(L_\theta = 8\pi)$, whereas in our simulations for $\eta=0.1$, intermittent patches of turbulence have not managed to self-organise into large-scale coherent structures.

\section{Discussion}

The numerical results from \S~3 and 4 have revealed three different onset regimes depending on the value of $\eta$ but also on the value of the numerical parameter $L_{\theta}$ :
\begin{enumerate}
\renewcommand{\labelenumi}{\roman{enumi})}
\item large-scale patterned turbulence at $\eta=0.8$, similar to all the planar regimes described in the literature
\item frustrated patterned turbulence at $\eta=0.5$, similar to the previous regime but emergent in simulations only when the azimuthal confinement is released 
\item a short-range laminar-turbulent coexistence regime at $\eta=0.1$, with no oblique patterning even for artificially large values of $L_{\theta}$, at least up to $16\pi$. 
\end{enumerate}
The first regime has a direct equivalent in aPf above $\eta \ge 0.4$ and thus does not deserve a very long description. It is notorious from several recent studies that this regime, near its onset, is heavily influenced by finite-size effects both in numerical simulations or in experiments. As for the long-standing question whether the evolution of $F_t$ with $Re$ is continuous or discontinuous, the answer for the patterned regime clearly follows the answer given for all Couette-like flows : in `affordable' numerical simulations, the transition at $Re_g$ appears discontinuous \citep{Bottin98,Duguet10,Chantry17}. However in the thermodynamic limit (infinite extent in space and time) it is continuous, which implies well-defined scalings for several quantities asymptotically close to onset. Two such scalings can be considered here from our data : a) the scaling of the turbulent fraction at equilibrium $F_t \sim \varepsilon^{\beta}$, where $\varepsilon$ is the normalised distance to the onset, i.e. $\varepsilon = (\Rey - Re_g)/Re_g$, and b) the scaling in time of the unsteady turbulent fraction during relaminarisation $F_t(t) \sim t^{-\alpha}$. The consensus at the moment is that this transition in shear flows is continuous. Besides it falls apparently into the universality class of directed percolation (DP), for which $\beta$ and $\alpha$ only depend on the effective dimension of the problem \citep{Lemoult16,Sano16,Chantry17}. In two spatial dimensions $\beta \approx 0.583$ and $\alpha \approx 0.451$,  whereas in one spatial dimension $\beta \approx 0.276$ and $\alpha \approx 0.159$. A continuous transition also implies the divergence of most correlations at the onset $Re_g$. This unfortunately implies that the determination of $Re_g$ itself, as well as that of exponents like $\beta$ and $\alpha$, also require divergent domain sizes and observation times, at the risk for the simulation or the experiment to become rapidly unfeasible. In practice however, the exponents are determined from finite-$\varepsilon$ simulations in finite domains over finite observation times, provided the domain length is several times the correlation length (and similarly with time). {In the patterning regime one can assume that the wavelength, for $\Rey$ away from $Re_g$, yields a decent estimate of the correlation length}; then the estimation of $Re_g$ requires a domain size at least one order of magnitude times this wavelength. Given that the patterning wavelength for $\eta=0.8$ and 0.5 is around $50h$ (cf figures~\ref{fig:tran08}b and \ref{fig:tran05_8pi}b), this suggests that no exponent can be trusted for sizes below $L_x \approx O(10^4)$, as confirmed in recent investigations of other shear flows \citep{Lemoult16,Chantry17}. Clearly this is beyond the present computational capacity for $\eta=0.8$ and 0.5.

\begin{figure}
\begin{center}
\includegraphics[height=6cm]{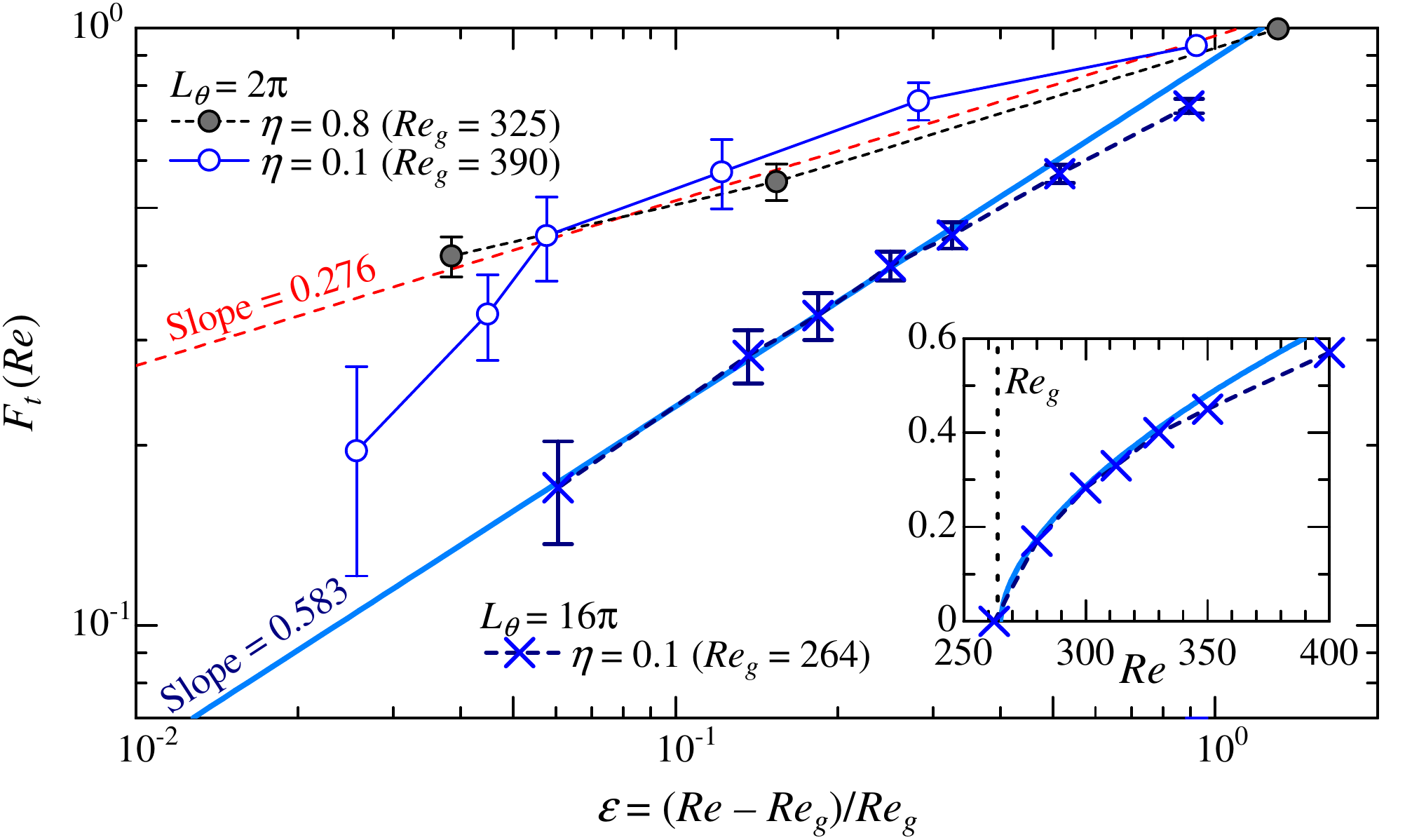}
\end{center}
\caption{Critical behaviour of the turbulent fraction as a function of the normalised distance $\varepsilon$ to the critical threshold $Re_g$ (whose value is specific to each case), in log-log scale. The two straight lines represent power law fits of the form $F_t \sim \varepsilon^{\beta}$ with $\beta_{\rm 2D} \approx 0.583$ as in (2+1)-DP (solid, blue online) and $\beta_{\rm 1D} \approx 0.276$ as in (1+1)-DP (dotted, red online). The inset shows the data for $\eta =0.1$ with $L_\theta = 16\pi$ in linear scale as in figure~\ref{fig:ft_all}.}
\label{fig:ft_epsilon}
\end{figure}

\begin{figure}
\begin{center}
\begin{tabular}{cc}
\begin{minipage}{0.49\textwidth}
\includegraphics[width=0.99\textwidth]{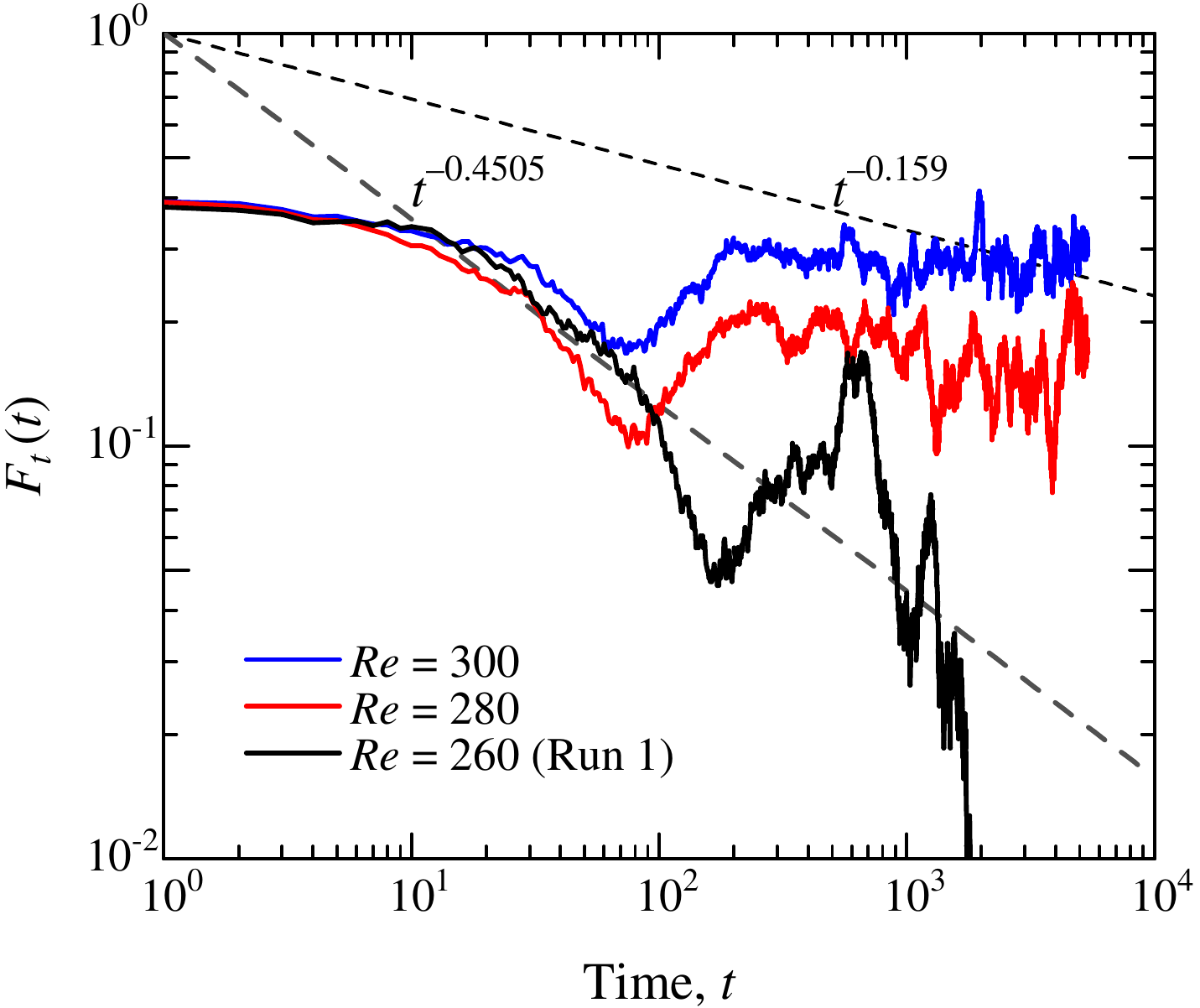}
\end{minipage}
\begin{minipage}{0.49\textwidth}
\includegraphics[width=0.99\textwidth]{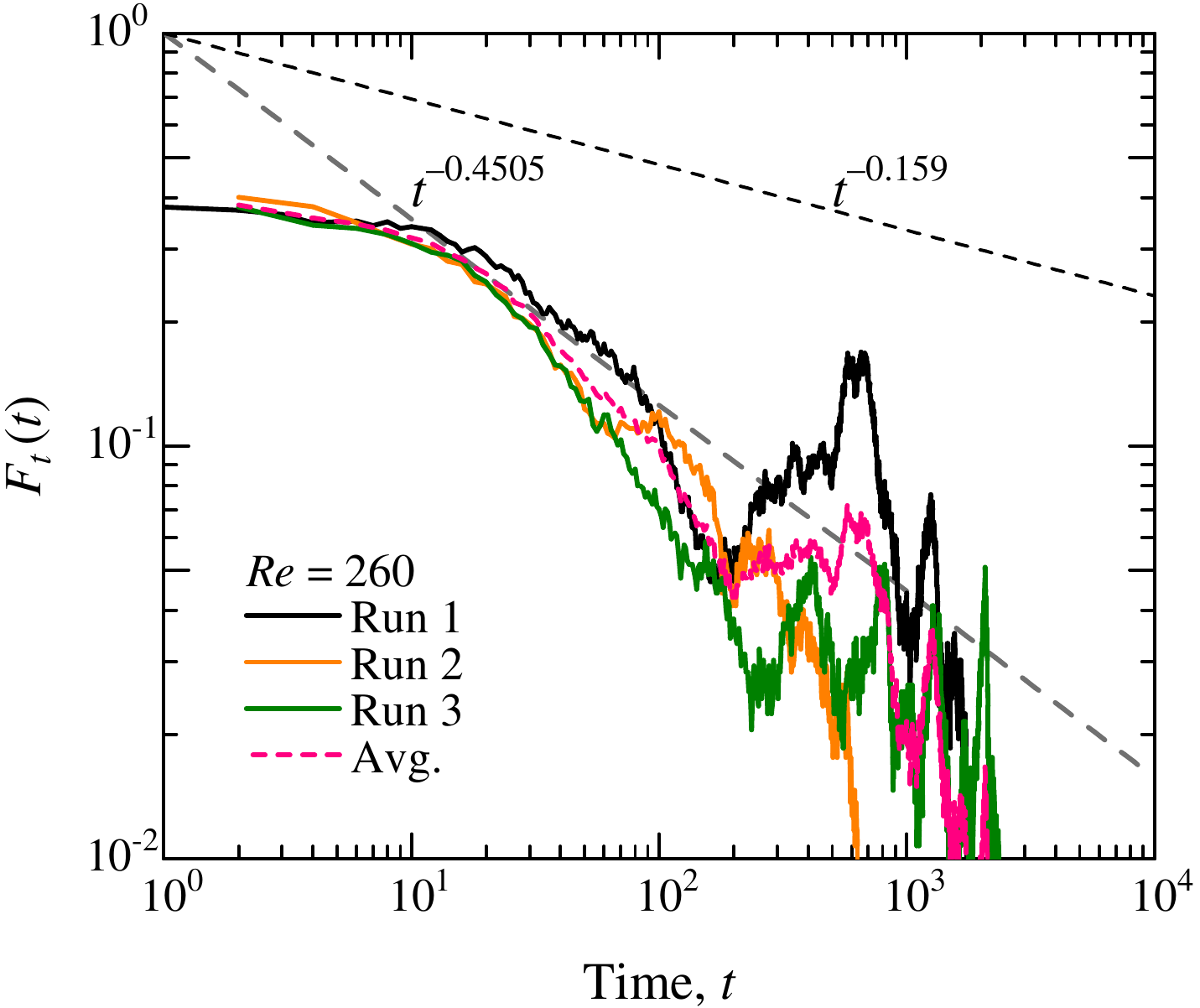}
\end{minipage}\\
\begin{minipage}{0.49\textwidth}
\centering
\vspace{0.5em}
(a) $\Rey = 260$--300
\end{minipage}
\begin{minipage}{0.49\textwidth}
\centering
(b) Three runs for $\Rey = 260$
\end{minipage}
\end{tabular}
\end{center}
\caption{Instantaneous turbulent fraction $F_t(t)$ as a function of time for $\eta = 0.1$ and $L_\theta = 16\pi$ near the critical Reynolds number. {Two dashed lines represent power laws of $F_t \propto t^{-\alpha}$ with the theoretical exponent either of $\alpha_{\rm 2D} = 0.4505$ for (2+1)-DP or $\alpha_{\rm 1D} = 0.159$ for (1+1)-DP.}}
\label{fig:ft_16pi}
\end{figure}

The discovery of a new regime for $\eta=0.1$, with short-range correlations in all directions, changes the picture. The short correlation lengths reported in \S~\ref{sec:tpc} suggest that, from the point of view of the turbulent fluctuations, the numerical domain with $L_x \approx 200h$ is more `extended', and in some sense closer to the thermodynamic limit of the related problem. We thus expect, for similar computational efforts as for other values of $\eta$, stronger evidence for continuous transition and easier measurements of the associated critical exponents. We thus replot the data for $\eta=0.1$ from figure~\ref{fig:ft_all} in log-log coordinates in order to verify whether the scaling  $F_t=O(\varepsilon^\beta)$ holds and, in case it does, which value $\beta$ takes. This is displayed in figure~\ref{fig:ft_epsilon} with $L_{\theta}=16\pi$, and $\varepsilon$ defined based on the choice {$Re_g=264$}. {The estimation of $Re_g$ is performed classically by trial and error, checking that the curve $F_t (Re)$ displays algebraic decay/growth for the fitted value of $Re=Re_g$.} A linear fit emerges over little less than a decade in $F_t$, which validates the notion of critical scaling. The fit for $\beta$ is consistent with the theoretical value $\beta_{\rm 2D}\approx 0.583$ for DP in two spatial dimensions (but not with the exponent $\beta_{\rm 1D}$). {Also plotted are the nominal cases of $L_\theta=2\pi$ for $\eta=0.1$ and 0.8, which exhibit respectively helically-shaped and puff-like turbulence. Both plots are consistent with $\beta_{\rm 1D}\approx 0.276$ rather than 0.583 (although a steep slope that deviates from $\beta_{\rm 1D}$ at $\varepsilon <0.06$ might be due to shortage in the streamwise domain length).} This alone is not sufficient to validate the {(2+1)-DP picture of the new regime} as two other independent exponents need to be validated as well. {Among these exponents}, we can have a rough approximation of the critical exponent $\alpha$ by monitoring in log-log plot the decay of $F_t(t)$ to laminar, starting from a noisy initial condition with finite $F_t$ at $t=0$. As can be seen in figure~\ref{fig:ft_16pi}a, the decay of an individual run for $\Rey \approx Re_g \approx 260$ alone cannot confirm the algebraic decay of $F_t$. There is always a possibility to get an improved critical scaling range by choosing more specific initial conditions with higher turbulent fraction. Ensemble averaging over several such runs is however helpful since an unambiguously algebraic decay for $F_t(t)$ emerges over {almost a decade} in figure~\ref{fig:ft_16pi}b. The measured exponent is then consistent with the theoretical value of $\alpha_{\rm 2D} \approx 0.45$. The current data appears hence consistent with (2+1)-DP, though more data and more exponents would be needed to properly confirm this trend. {Note in figure~\ref{fig:ft_16pi} that the scaling regime is reached after a relatively short time of $O(10h/u_w)$, especially compared with other similar studies \citep[e.g.,][]{Lemoult16,Chantry17}. This is consistent in order of magnitude with the shorter decay and splitting times observed in animations. This is another major advantage for gathering equivalent statistics compared to other flows displaying continuous transition.} 

{By contrast}, for the other cases $\eta=0.8$ and 0.5 reported here, the numerical domain is so confined that no trend emerges : whatever the thermodynamic limit might be, the transition in that case only appears in practice as discontinuous. In other words, we have uncovered for $\eta$=0.1 a new dynamical regime with short-range correlations, for which verification of the DP property seems computationally feasible using realistic domain sizes. This is in marked contrast to the majority of planar subcritical shear flows in which a similar task would still today appear as computationally hopeless.


\section{Conclusion}

Direct numerical simulation of annular Couette flow (aCf) {is reported} using finite differences in long computational domains. The so-called transitional regimes of aCf, featuring coexistence of laminar and turbulent flow, {have been} investigated depending on the radius ratio $\eta$. The influence of an additional numerical parameter, 
 the {azimuthal} $L_\theta$ (usually fixed to $2\pi$), {has also been} considered as a way to question the direct influence of azimuthal confinement on the possible formation of large-scale flows
 and thus of organised laminar-turbulent coexistence. Three different regimes {have been} identified. For $\eta$ close to unity (e.g. here 0.8) large-scale helical bands form as in the planar limit of plane Couette flow (pCf).
 For moderate $\eta$ (e.g. here 0.5), these helical bands do not have enough space to form because of the azimuthal confinement, this is confirmed by their occurrence for $L_{\theta}$ sufficiently large with respect to $2\pi$.
 Eventually, for low enough $\eta$, turbulence near its onset takes the form of disorganised patches; their localisation and interaction {are} reminiscent of turbulent puffs in pipe flow but their structure displays a strong asymmetry. Importantly, the correlation length between these puffs is shorter by one order of magnitude than the coherence length of organised patterns. This new regime appears as a potential candidate for directed percolation, the short-range property even suggests that critical exponents could be measured with significantly less computational effort than for the other long-range regimes usually encountered in subcritical shear flows. It remains an open question whether it is possible to identify other shear flow geometries in which turbulent patches would also display short-range correlations, preferably a realistic flow that can be released experimentally.
 
 {In this study, aCf has been introduced as a continuation prototype linking pCf to a one-dimensional flow geometry, with cylindrical pipe flow as the canonical example for a one-dimensional geometry. The present results indicate that high-$\eta$ aCf connects smoothly with pCf, however it fails at connecting with the pipe flow for low $\eta$ and is thus no relevant candidate for this continuation. The presence of the inner rod, together with the no-slip condition on it, induces a mean flow differing strongly from the one expected for a pipe flow}. In particular, the statistics for $\eta=0.1$ clearly demonstrate that, as for the laminar case, the highest shear is found near the inner rod, whereas the shear at the outer wall gets comparatively weaker as $\eta$ is reduced. This corresponds to the opposite situation to the pipe flow, where the shear is expected to vanish near the axis. The strong shear near the {inner} rod is responsible for the occurrence of a new dynamical regime characterised by short-range correlations and no large-scale organisation.

\section*{Acknowledgements}

This work was supported by Grant-in-Aid for JSPS (Japan Society for the Promotion of Science) Fellowship 16H06066, 16H00813, and 19H02071. Numerical simulations were performed on SX-ACE supercomputers at the Cybermedia Centre of Osaka University and the Cyberscience Centre of Tohoku University.

\end{document}